\documentclass[twocolumn,floats,floatfix,
               showpacs,prd,superscriptaddress,
               longbibliography,
               nofootinbib]{revtex4-2}

\usepackage[utf8]{inputenc}
\usepackage{graphicx,epsfig}
\usepackage{epstopdf}
\usepackage{amssymb,amsmath,amsthm,amsfonts}
\usepackage{bm}
\usepackage[caption=false]{subfig}
\usepackage[usenames,dvipsnames]{xcolor}
\usepackage{url}
\usepackage[inline]{enumitem}
\usepackage{dsfont}
\usepackage{float}
\usepackage{placeins}
\usepackage[T1]{fontenc}
\usepackage{mathrsfs}
\usepackage{anyfontsize}
\usepackage{xspace}
\usepackage{tensor}
\usepackage{microtype}
\usepackage{tikz}
\usetikzlibrary{arrows.meta}
\usepackage{longtable}
\usetikzlibrary{positioning}
\usepackage{acronym}
\usepackage{comment}
\usepackage[linktocpage,breaklinks]{hyperref}
\hypersetup{colorlinks=true,
            citecolor=NavyBlue,
            linkcolor=NavyBlue,
            urlcolor=NavyBlue}

\renewcommand{\vec}[1]{\boldsymbol{#1}}

\def\newacronym#1#2#3{\gdef#1{\gdef#1{#2\xspace}#3 (#2)\xspace}}
\def\bh#1{black hole#1 (BH#1)\gdef\bh{BH}}
\def\bbh#1{binary black hole#1 (BBH#1)\gdef\bbh{BBH}}
\def\gw#1{gravitational wave#1 (GW#1)\gdef\gw{GW}}
\newacronym{\bssn}{BSSN}{Baumgarte-Shapiro-Shibata-Nakamura}
\newacronym{\CAH}{CAH}{common apparent horizon}

\newacronym{\ode}{ODE}{ordinary differential equation}
\newacronym{\pde}{PDE}{partial differential equation}

\newacronym{\eft}{EFT}{effective field theory}
\newacronym{\eob}{EOB}{effective-one-body}

\newacronym{\hpc}{HPC}{High-performance computing}

\def\canuda{\textsc{Canuda}}
\def\ETK{\textsc{Einstein Toolkit}}

\def\dif{\textrm{d}}

\def\R4D{R}
\def\rex{r_{\rm ex}}

\begin{document}

\title{Spin-up and mass-gain in hyperbolic encounters of spinning black holes}

\author{Healey Kogan}
\email{hjkogan@stanford.edu}
\affiliation{Department of Physics, Stanford University, Stanford, California 94305, USA}
\affiliation{The Grainger College of Engineering,
Department of Physics \& Illinois Center for Advanced Studies of the Universe, University of Illinois Urbana-Champaign, Urbana, Illinois 61801, USA}

\author{Frederick C.L. Pardoe}
\email{fpardoe2@illinois.edu}
\affiliation{The Grainger College of Engineering,
Department of Physics \& Illinois Center for Advanced Studies of the Universe, University of Illinois Urbana-Champaign, Urbana, Illinois 61801, USA}

\author{Helvi Witek}
\email{hwitek@illinois.edu}
\affiliation{The Grainger College of Engineering,
Department of Physics \& Illinois Center for Advanced Studies of the Universe, University of Illinois Urbana-Champaign, Urbana, Illinois 61801, USA}
\affiliation{Center for AstroPhysical Surveys, National Center for Supercomputing Applications, University of Illinois Urbana-Champaign, Urbana, IL, 61801, USA}

\begin{abstract}
Scattering black holes 
spin up and gain mass through the re-absorption of orbital angular momentum and energy radiated in gravitational waves during their encounter. 
In this work, we perform a series of  numerical relativity simulations to investigate the spin-up and mass-gain
for equal-mass black holes 
in a wide range of equal initial spins, $\chi_{\rm i}\in[-0.7,0.7]$, aligned (or anti-aligned) to the orbital angular momentum. 
We also consider a variety of initial momenta. Furthermore, we explore a range of incident angles and identify the threshold between scattering and merging configurations.
The spin-up and mass-gain are typically largest in systems with incident angles close to the threshold value, large momenta, and negative (i.e. anti-aligned) initial spins.
When evaluated at the threshold angle, we find that the spin-up decreases linearly with initial spin. 
Intriguingly, systems with initial spin $\chi_{\rm i}=0.7$ sometimes experience a spin-down, in spite of an increase in the black-hole angular momentum, due to a corresponding gain in the black-hole mass. 
Across the simulation suite, we find a maximum spin-up of $0.3$ and a maximum increase in the black-hole mass of $15\%$.
\end{abstract}

\maketitle

\section{Introduction} \label{Sec:Introduction}

Historically, most efforts devoted to the study of binary \bh{s} have focused on quasi-circular coalescences. 
In these systems, the \bh{s} begin gravitationally bound on approximately circular orbits, which decay through \gw{} emission until the \bh{s} merge. 
The rationale behind this emphasis is that binaries circularize as a consequence of \gw{} emission~\cite{Peters:1964zz,Hinder:2007qu} and thus ought to be quasi-circular by the time they enter the LIGO band.
Since the first detection of \gw{s} in 2015~\cite{LIGOScientific:2016aoc}, this reasoning has been largely validated through observations~\cite{LIGOScientific:2025slb}. 
However, it is thought that a small number of eccentric mergers may have been detected, likely resulting from dynamical capture in hyperbolic systems~\cite{Gamba:2021gap, Gayathri:2020coq}.

Hyperbolic binaries fall into three morphological categories: 
they either merge, scatter, or undergo a zoom-whirl, where the \bh{s} perform a series of small and large orbits prior to merging.
The physics of hyperbolic encounters (i.e. the interaction of scattering \bh{s}) is particularly interesting in light of upcoming \gw{} experiments. 
These encounters are expected to be found by third generation detectors~\cite{Mukherjee:2020hnm,Garcia-Bellido:2021jlq,Kerachian:2023gsa}, 
such as the Cosmic Explorer~\cite{Reitze:2019iox,Evans:2021gyd},
Einstein Telescope~\cite{Punturo:2010zz,ET:2025xjr},
and LISA~\cite{LISAConsortiumWaveformWorkingGroup:2023arg,LISA:2022yao}.
They may even be detectable with current ground-based instruments~\cite{
Kocsis:2006hq, Mukherjee:2020hnm,LIGOScientific:2014pky,Morras:2021atg,Bini:2023gaj}, given improved data analysis.

Hyperbolic encounters are thought to be common in dense clusters~\cite{Sigurdsson:1993zrm,Lightman:1978zz,Rodriguez:2019huv}.
Clusters are typically modeled using N-body simulations in (post-) Newtonian gravity (e.g.~\cite{Joshi:1999vf,Pattabiraman:2012ti,Rodriguez:2019huv,Gaete:2024ovu,Siles:2024yym,Kiroglu:2024xpc}), although there have been advancements in the use of full general relativity~\cite{Bamber:2025gxj}.
Studying hyperbolic encounters can help to improve these models, which in turn can explain the astrophysical origin of the encounters.
This combined knowledge then contributes to
the understanding of \bh{} formation channels, primordial \bh{s}, \gw{} sources, and other astrophysical phenomena.

Early numerical studies on \bh{} binaries outside of the quasi-circular regime identified and analyzed the morphologies described above~\cite{Pretorius:2007jn,Shibata:2008rq,Sperhake:2009jz,Healy:2009zm,Gold:2009hr,Gold:2012tk}.
The impact of the different morphologies on the gravitational radiation has been computed in conjunction with (semi-) analytic methods~\cite{Berti:2010ce,East:2012xq,Damour:2014afa,Nagar:2020xsk,Rettegno:2023ghr}.
The \gw{} emissions from the encounters have been computed in a series of works~\cite{Garcia-Bellido:2017qal,Teuscher:2024xft,Caldarola:2023ipo,Roskill:2023bmd,Bae:2023sww,Fontbute:2024amb}.

In addition to applications in \gw{} astrophysics, simulations of \bh{s} have also been used to model high-energy particle collisions.
These simulations focused on ultra-relativistic \bh{} scattering or mergers~\cite{Shibata:2008rq,Sperhake:2009jz,Sperhake:2010uv,Sperhake:2012me},
and it was found that with increasing initial momentum, 
the binaries' morphology and radiated energy
became less sensitive
to internal parameters like the \bh{s'} spins.

Another line of research has investigated the deflection angle imparted on the trajectories of scattering \bh{s}~\cite{Damour:2014afa,Hopper:2022rwo,Rettegno:2023ghr,Long:2025nmj},
connecting to new \gw{} modeling methods such as scattering amplitudes. Recent numerical work has striven toward the creation of \gw{} catalogs for eccentric mergers~\cite{Trenado:2025ccf,Ficarra:2024nro,Nee:2025zdy}. 

In this work, we focus on the evolution of the spins and masses of scattering \bh{s}. Simulations of ultra-relativistic scattering \bh{s} revealed that their (dimensionless) spins increase along the \bh{} angular momentum if they are initially zero or anti-aligned to it ~\cite{Sperhake:2012me}.
This phenomenon, also called the ``spin-up,'' 
occurs due to the transfer of orbital angular momentum to the \bh{s} via the re-absorption of \gw{s}.
It was also shown that initially aligned \bh{} spins could decrease (i.e. ``spin-down'').

An analysis of scattering equal-mass, non-spinning \bh{s} with moderate initial momenta showed that the spin-up becomes more pronounced for small incident angles and large initial momenta~\cite{Nelson:2019czq}.
In binaries of unequal-mass \bh{s}, the more massive \bh{} undergoes a larger spin-up~\cite{Jaraba:2021ces}.
Simulations of precessing \bh{s} have shown that the spin-up decreases with increasing initial spins aligned to the orbital angular momentum and increases with initial spins that are orthogonal to the orbital angular momentum~\cite{Rodriguez-Monteverde:2024tnt}.
Complementary work employing the effective-one-body approach has enabled the modeling of the spin evolution in dense clusters~\cite{Nagar:2020xsk,Chiaramello:2024unv}. 
The broader phenomenon of spin change due to the re-absorption of angular momentum emitted in gravitational radiation is called ``tidal-torquing,''
and it has been explored using (semi-) analytic models~\cite{Tagoshi:1997jy,Alvi:2001mx,Poisson:2004cw,Taylor:2008xy,Comeau:2009bz,Poisson:2009qj,Chatziioannou:2012gq,Poisson:2014gka,Chatziioannou:2016kem,Poisson:2018qqd,Saketh:2022xjb,Chiaramello:2024unv,Zi:2023geb,Mukherjee:2025wxa}. 
These studies also consider a sister phenomenon called ``tidal-heating,'' whereby \bh{s} gain mass from the re-absorption of energy emitted in gravitational radiation.

The present study has two principal goals: 
the first is to further explore the spin-up of scattering \bh{s} in new regions of parameter space,
and the second is to analyze their mass-gain.
In this work, we consider initially spinning, equal-mass \bh{s} with (dimensionless) spins in the range $\chi\in [-0.7,0.7]$ for a series of moderate initial momenta and incident angles.
This spin range is particularly interesting for \bh{s} formed in previous mergers
as merger remnants typically have spins around $\chi\sim0.7$~\cite{Gammie:2003qi,Sperhake:2007gu,Hofmann:2016yih,Rodriguez-Monteverde:2025rfh}. \bh{s} in dense clusters have also been shown to acquire spins in the range $\chi\sim0.4-0.9$~\cite{Rodriguez:2019huv,Bamber:2025gxj}. 
Furthermore, there have been detections of mergers in which one of the \bh{} spins was initially negative \cite{LIGOScientific:2025brd}.
Moreover, it is thought that repeated spin-ups may play a key role in determining the spin distribution of primordial \bh{s} \cite{Garcia-Bellido:2020pwq,Jaraba:2021ces,Siles:2024yym,Rodriguez-Monteverde:2024tnt}.

We structure this work as follows. In Sec.~\ref{sec:Setup}, we describe the setup, computational details,
simulation suite and validation tests.
The results are shown in Sec.~\ref{Sec:Results}. Specifically, Sec.~\ref{Sec:Results_Morphology} presents the systems' morphologies, while
Secs.~\ref{Sec:Results_Spinup} and~\ref{Sec:Results_Mass}
present the spin-up and mass-gain observed in scattering systems.
In Sec.~\ref{Sec:Conclusion}, we discuss our conclusions.
Appendix~\ref{sec:convergence_tests}
provides a detailed analysis of
convergence tests and uncertainties.
Throughout, we use geometric units $G=1=c$.

\section{Setup and Numerical framework} \label{sec:Setup}

\subsection{Initial configuration of black hole binary} \label{sec:Setup_Configuration}

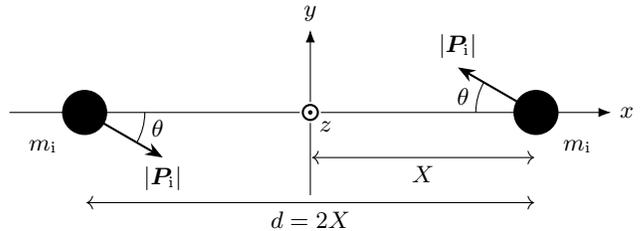
\begin{figure}[htbp!]
    \begin{center}
  \begin{tikzpicture}
    \fill[black] (-3 , 0) circle (0.3 ) node[below left=0.35] {$m_{\rm i}$};
    \draw[thick,-Stealth] (-3,0) -- (-3+0.866*1.2,-0.5*1.2) node[below] {$|\vec{P}_{\rm i}|$};
    \draw (-3+0.8,0) arc (0:-30:0.8) node[midway, right] {$\theta$};

    \fill[black] (3, 0) circle (0.3 ) node[below right=0.35] {$m_{\rm i}$};
    \draw[thick,-Stealth] (3,0) -- (3-0.866*1.2,0.5*1.2) node[above] {$|\vec{P}_{\rm i}|$};
    \draw (3-0.8,0) arc (180:150:0.8) node[midway, left] {$\theta$};

    \draw[-Latex] (-4,0) -- (4,0) node[right] {$x$};
    \draw[-Latex] (0,-1.1) -- (0,1.1) node[above] {$y$};
    \fill[white] (0 , 0) circle (0.14 );
    \draw[thick] (0 , 0) circle (0.1 ) node[below right] {$z$};
    \fill[black] (0 , 0) circle (0.03 );

    \draw[<->] (-2.975,-1.2) -- (2.975,-1.2) node[midway,below] {$d=2X$};
    \draw[<->] (0.025,-0.6) -- (2.975,-0.6) node[midway,below] {$X$};

  \end{tikzpicture}
    \caption{\label{fig:InitialConditions} Initial conditions of binary \bh{s} with equal initial masses, $m_{\rm i}$, total mass $\mathrm{M}=2m_{\rm i}$, and equal initial spins, $\chi_{\rm i}$. The spins are aligned or anti-aligned with the orbital angular momentum, which points in the z-direction.
    The setup has rotational symmetry such that the \bh{s} have equal but opposite initial (linear) momenta, $|\vec{P}_{\rm i}|$, inclined at an incident angle, $\theta$, from the x-axis. Furthermore, the \bh{s} have an initial separation $d=100\mathrm{M}$ along the x-axis.
    }
    \end{center}
\end{figure}

In this work, we consider the scattering or merger of two \bh{s} with equal masses, $m$, and equal (dimensionless) spins, $\chi=S/m^2$, where $S$ refers to the \bh{} angular momentum.
Their initial setup is depicted in Fig.~\ref{fig:InitialConditions}.
The \bh{s} are initially located along the x-axis,
each at a distance $X$ from the origin so that their initial separation is $d=2X$.
The \bh{s} have equal, but oppositely directed, initial (linear) momenta, $|\vec{P}_{\rm i}|$, with an incident angle, $\theta$, with respect to the x-axis. The \bh{s}' initial spins are aligned or anti-aligned with the orbital angular momentum (i.e. along the z-axis). We refer to the initial values of the \bh{} mass and spin as $m_{\rm i}$ and $\chi_{\rm i}$, respectively.
The total mass of the system is $\mathrm{M}=2m_{\rm i}=1$ (in code units).

Here and henceforth, the \bh{} mass, $m$, refers to the Christodoulou mass. The (dimensionless) spin, $\chi \in (-1,+1)$, is positive when aligned with the orbital angular momentum and negative when anti-aligned with the orbital angular momentum. 
The linear momentum $P$ 
and \bh{} mass $m$ are initially prescribed 
by the Bowen-York linear momentum and the individual \bh{s'} ADM mass~\cite{Bowen:1980yu,Brandt:1997tf}.
We note that the \bh{} mass coincides with the Christodoulou mass in the limit of isolated \bh{s}.

Holding other parameters constant, as the incident angle is decreased, the \bh{s} scatter, follow zoom-whirl orbits, or merge; see Fig.~\ref{fig:Boost245Spin00}. The lowest incident angle for which a system undergoes scattering is called the threshold angle, $\theta_{\rm th}$. We use the subscripts ``$1$'' and ``$2$'' to refer to the initial \bh{s}. In cases where the \bh{s} merge, we use the subscript ``$3$'' to refer to the remnant. All dimensionful quantities are expressed in units of the total mass, $\mathrm{M}=1$.

\subsection{Extraction of observables} \label{sec:Setup_Observables}

We require a variety of information about the properties of the \bh{s} and the \gw{s} that they produce. In particular, we seek to understand how the \bh{s}' masses and spins change as a consequence of scattering. To analyze the \bh{s}' evolution, we extract the Weyl scalar, $\Psi_{4}$, and compute properties of the \bh{s'} apparent horizons.

The Weyl scalar is a measure of the outgoing gravitational radiation. It is decomposed into multipoles by projecting it onto spin-weighted spherical harmonics,
\begin{align}
\label{eq:Weyl_Decomp}
\Psi_{4,\mathfrak{lm}} (t,\rex) & =
    \int \dif\Omega\, \Psi_{4}(t,\rex,\theta,\phi) Y^{\ast}_{-2,\mathfrak{lm}}(\theta,\phi)
\,,
\end{align}
where $Y^*_{-2,\mathfrak{lm}}(\theta,\phi)$ are the complex conjugates of the spin-weighted spherical harmonics, $Y_{\mathfrak{s,lm}}(\theta,\phi)$, with spin-weight $\mathfrak{s}=-2$.
The integration in the formula is carried out over a sphere of extraction radius, $\rex$.

We are particularly interested in \bh{} properties. 
The simulation output includes the horizon area, $A_{\rm H}$, the irreducible mass, $m_{\rm irr}=\sqrt{A_{\rm H}/16 \pi}$, and the equatorial circumference, $C_{e}$, 
of the apparent horizon in the plane perpendicular to the \bh{} spin,
among other quantities.
We use them to compute the spin and \bh{} angular momentum~\cite{Kiuchi:2009jt,Witek:2010qc},
\begin{subequations}
\label{eq:extract}
\begin{align}
\label{eq:extract_spin}
& \chi = \sqrt{1-\left( \frac{2 \pi A_{\rm H}}{C_{\rm e}^2}-1 \right)^2}\,,\\
\label{eq:extract_angmom}
& S=m^2 \chi  \,.
\end{align}
\end{subequations}
We compute the \bh{} mass from the equatorial circumference of the horizon~\cite{Kiuchi:2009jt}; 
however, it is physically insightful to recall that it can also be expressed in terms of the previous quantities using the Christodoulou formula~\cite{Christodoulou:1970wf,Kiuchi:2009jt,Gerosa:2022fbk},
\begin{subequations}
\label{eq:mchris}
\begin{align}
& m_{\rm} = \frac{C_{\rm e}}{4 \pi} \,,
\label{eq:mchris_extract} \\
& m^2 = m^2_{\rm irr} + \frac{S^2}{4 m^2_{\rm irr} } = \frac{2 m_{\rm irr}^2}{1+\sqrt{1-\chi^2}} \,.
\label{eq:mchris_angmom}
\end{align}
\end{subequations}
We see that the \bh{} mass, or Christodoulou mass, originates from the sum of the irreducible mass and the \bh{} angular momentum. In particular, Eq.~\eqref{eq:mchris_angmom} implies that the \bh{} mass can change when the \bh{} spin changes, even if the irreducible mass remains the same. We remark upon this consequence further when discussing how these quantities change as the result of scattering. 
Eqs.~\eqref{eq:extract} and~\eqref{eq:mchris} only apply to isolated \bh{s}, but we can use them to study \bh{} binaries so long as the \bh{s} are widely separated, or after their remnant has settled, should the \bh{s} merge.
In using these formulae to quantify the \bh{} mass and spin, we follow Ref.~\cite{Sperhake:2012me}. 
Other methods for computing the spin exist, such as via a ratio of the apparent horizon's circumferences or quasi-local measures, which give comparable results; see e.g. Refs.~\cite{Anninos:1994pa,Kiuchi:2009jt,Nelson:2019czq}.

In the simulations, we find that the \bh{s} typically spin up while scattering, and that the increase in spin is largest near the scattering threshold. This behavior is due to the re-absorption of energy and orbital angular momentum
emitted in \gw{s}.
In order to understand the role that the orbital angular momentum, $J$, plays in the evolution of the \bh{s'} spins,
we compute its values before and after the close encounter in scattering simulations.
Note that the orbital angular momentum only has a $z$-component due to the symmetries of the system. Therefore, we exclusively refer to this component, rather than to the whole vector. We compute the initial orbital angular momentum, $J_{\rm i}$, from the initial (linear) momentum,
\begin{equation}
\label{eq:Orbital_AngMom_initial}
J_{\rm i} =2X|\vec{P}_{\rm i}|\sin{\theta}
\,,
\end{equation}
where $d=2X$ is the \bh{s'} initial separation and $\theta$ is the incident angle.
To compute the final orbital angular momentum, $J_{\rm f}$, we follow Refs.~\cite{Nelson:2019czq,Rodriguez-Monteverde:2024tnt} and utilize global conservation of the angular momentum. Thus,
\begin{equation}
\label{eq:Orbital_AngMom_final}
J_{\rm f} = J_{\rm i}-J_{\rm GW}-2(S_{\rm f}-S_{\rm i})
\,,
\end{equation}
where $S_{\rm i}$ and $S_{\rm f}$ are the initial and final \bh{} angular momentum before and after scattering, respectively. $J_{\rm GW}$ is the angular momentum radiated away by \gw{s}, which we compute
as~\cite{Campanelli:1998jv,Nelson:2019czq},
\begin{equation}
\label{eq:Orbital_AngMom_GW_1}
J_{\rm GW} = \frac{\rex^2}{16 \pi} \sum_{\mathfrak{l,m}} \int -\mathfrak{m}(\dot{h}^{+}_{\mathfrak{lm}}h^{\times}_{\mathfrak{lm}}-\dot{h}^{\times}_{\mathfrak{lm}}h^{+}_{\mathfrak{lm}})dt \,.
\end{equation}
Here, $h^{+}$ and $h^\times$ are polarizations of the \gw{} strain,
and a dot denotes derivatives with respect to time.
They are related to the Weyl scalar via,
\begin{equation}
\label{eq:Orbital_AngMom_GW_2}
\Psi_{\mathrm{4,}\mathfrak{lm}} = -\ddot{h}^{+}_{\mathfrak{lm}}+i\ddot{h}^{\times}_{\mathfrak{lm}} \,,
\end{equation}
where the separation of the Weyl scalar into its real and imaginary components is given in Ref.~\cite{Alcubierre:2008}. To find the radiated angular momentum, we first integrate over these components to find the strain polarizations and their derivatives as functions of time. We then integrate Eq.~\eqref{eq:Orbital_AngMom_GW_1} from the simulation start time until the time at which the radiation emitted from the encounter passes through the extraction radius.
In practice, we use an extraction radius of $\rex=100\mathrm{M}$ and sum over $\mathfrak{l}\in[0,6]$, $\mathfrak{m}\in[-\mathfrak{l},\mathfrak{l}]$.

\subsection{Code description} \label{sec:Setup_Code}

In this work we conduct simulations with the \ETK~\cite{maxwell_rizzo_2025_15520463,Loffler:2011ay, Zilhao:2013hia}, an open-source software for computational astrophysics, and the \canuda~code~\cite{witek_2023_7791842,Okawa:2014nda,Zilhao:2015tya,Witek:2018dmd,Silva:2020omi,Richards:2025ows} for fundamental physics.
The \ETK~is built upon the \textsc{Cactus} computational framework \cite{Goodale2002a,Cactuscode} and uses \textsc{Carpet} \cite{Schnetter:2003rb,Carpetbitbucket} to implement box-in-box adaptive mesh refinement as well as hybridized message passing interface and open multi-processing parallelization.

This software evolves \bh{} binaries using a 3+1 formulation of Einstein's equations, where the four dimensional spacetime is foliated into a series of three dimensional hypersurfaces parameterized by the time, $t$.
Given initial data for the induced metric on a hypersurface and its extrinsic curvature,
the evolution equations are solved using the method of lines.

To generate initial data, we use the \verb|TwoPunctures| spectral thorn \cite{Ansorg:2004ds}, which solves the constraint equations via the Bowen-York method as extended by Brandt and Brügmann~\cite{Bowen:1980yu,Brandt:1997tf}.
We then evolve the system using \canuda's \verb|LeanBSSNMoL|\footnote{This thorn is adapted from the \texttt{Lean} code \cite{Sperhake:2006cy}.} thorn, which implements the \bssn formalism \cite{Shibata:1995we,Baumgarte:1998te} together with the moving puncture gauge \cite{Campanelli:2005dd,Baker:2005vv}.
\verb|LeanBSSNMoL| provides up to eighth order finite differences for spatial derivatives.
Here, we use fourth order finite differencing for spatial derivatives and employ the fourth order Runge-Kutta scheme for the time integration.

We obtain data on the gravitational radiation by computing the Weyl scalar, $\Psi_{4}$, with \canuda's \verb|NPScalars| thorn.
We then use the \verb|Multipole| thorn~\cite{MultipoleThorn} to
project the Weyl scalar into its multipoles
$\Psi_{4,\mathfrak{lm}}$
using Eq.~\eqref{eq:Weyl_Decomp}.
These modes are computed on  spheres of constant extraction radii, $\rex$.
We compute the \bh{} apparent horizons and their properties using the \verb|AHFinderDirect| thorn \cite{Thornburg:1995cp,Thornburg:2003sf}.

\subsection{Summary of simulation suite} \label{sec:Setup_Simulation_Suite}

To investigate the effect of initial spin on the evolution of \bh{} binaries, we perform an extensive simulation suite that is summarized in Table~\ref{tab:RunsScatteringBoostP0245P0490}.
The suite consists of about $150$ simulations and we estimate their total computational cost to be of the order of $10^6$ core hours. 
In each simulation, the \bh{s} have an initial separation $d=100\mathrm{M}$ along the x-axis.
We use equal-mass \bh{s} with initial mass $m_{\rm i}=0.5$, such that we have an initial total mass $\mathrm{M}=1$ in code units.
In the first simulation suite, we vary the initial spins (equal for each \bh{})
in the range
$\chi_{\rm i}=\{-0.7,\, -0.5,\, -0.2,\, 0.0,\, 0.2,\, 0.5,\, 0.7\}$;
a positive (negative) sign corresponds to an initial spin that is aligned (anti-aligned) with the orbital angular momentum.
We consider initial (linear) momenta $|\vec{P}_{\rm i}|/\mathrm{M}=\{0.245,\,0.490\}$.

\begin{table}[htpb!]
\begin{tabular}{|c|c|c|c|c|}
\hline
 Series & $|\vec{P}_{\rm i}|/\mathrm{M}$ & $\chi_{\rm i}$ & $\theta_{\rm M}$
       & $\theta_{\rm S}$ \\
\hline
Xm7P24 & $0.245$   &  $-0.7$  &  $(0.04700,0.06725)$   & $(0.06750,0.07200)$ \\
Xm5P24 & $0.245$   &  $-0.5$  &  $(0.06300,0.06500)$   & $(0.06525,0.06700)$ \\
Xm2P24 & $0.245$   &  $-0.2$  &  $(0.05500,0.06100)$   & $(0.06125,0.06500)$ \\
Xp0P24 & $0.245$   &  $0.0$  &  $(0.05700,0.05800)$   & $(0.05825,0.06500)$ \\
Xp2P24 & $0.245$   &  $0.2$  &  $(0.05200,0.05500)$   & $(0.05525,0.06500 )$ \\
Xp5P24 & $0.245$   &  $0.5$  &  $(0.04800, 0.05050)$   & $(0.05075,0.05500)$ \\
Xp7P24 & $0.245$   &  $0.7$  &  $(0.04500, 0.04800)$   & $(0.04825, 0.05400)$ \\
\hline
Xm7P49 & $0.490$   &  $-0.7$  &  $(0.05600,0.05680)$   & $(0.05685,0.05780)$ \\
Xm5P49 & $0.490$   &  $-0.5$  &  $(0.05200, 0.05450)$   & $(0.05500, 0.05600)$ \\
Xm2P49 & $0.490$   &  $-0.2$  &  $(0.04500, 0.05200)$   & $(0.05250, 0.05500)$ \\
Xp0P49 & $0.490$   &  $0.0$  &  $(0.04800,0.04975)$   & $(0.05000,0.06000)$ \\
Xp2P49 & $0.490$   &  $0.2$  &  $(0.04600, 0.04775)$   & $(0.04800, 0.05400 )$ \\
Xp5P49 & $0.490$   &  $0.5$  &  $(0.04000, 0.04400)$   & $(0.04500, 0.05000)$ \\
Xp7P49 & $0.490$   &  $0.7$  &  $(0.04000, 0.04200)$   & $(0.04250, 0.05685)$ \\
\hline
\end{tabular}
\caption{\label{tab:RunsScatteringBoostP0245P0490}
We summarize a set of \bh{} binary simulations with initial (linear) momenta $|\vec{P}_{\rm i}|/\mathrm{M}=\{0.245, 0.490\}$, initial spins $\chi_{\rm i}\in[-0.7,0.7]$, and the range of incident angles resulting in either merger $\theta_{\rm M}$, or scattering $\theta_{\rm S}$. Negative spin indicates anti-alignment with the orbital angular momentum.}
\end{table}

For each combination of initial spin and initial momentum, we run a set of simulations with varying incident angles.
We seek to find at least one angle that results in a merger and explore a sufficient range of the scattering parameter space such that we can comment on qualitative changes as a function of the incident angle.
To identify the angle that indicates the threshold between
the scattering and the merger of \bh{s},
we start with the results of Ref.~\cite{Nelson:2019czq} for vanishing initial spin, $\chi_{\rm i}=0$.
Then, we typically vary the angle in intervals of $1\times10^{-3}$. Once we find the boundary between the merging and scattering simulations, we further explore the parameter space between them by iterating over typical differences of $2.5\times10^{-4}$ until we find the boundary again. We then refer to the smallest angle which results in scattering as the threshold angle, $\theta_{\rm th}$.
However, the ``true'' threshold is between the largest incident angle resulting in a merger, and the smallest incident angle resulting in a scattering.
We take the spacing between them, $\sim 2.5\times10^{-4}$, as a conservative estimate of the error in the threshold angle.
In Table~\ref{tab:RunsScatteringBoostP0245P0490}, we indicate the range of incident angles which result in a merger as $\theta_{\rm M}$,
and those that result in a scattering as $\theta_{\rm S}$.

In systems with initial spin $\chi_{\rm i}=0.7$, we notice qualitatively different trends in the change in spin, which depend on the initial momentum.
To further explore these trends, we run a second simulation suite,
summarized in Table~\ref{tab:RunsScatteringSpin0P7},
with fixed initial spins $\chi_{\rm i}=0.7$ and varying initial momenta $|\vec{P}_{\rm i}|/\mathrm{M}=\{0.06125,\, 0.1225,\, 0.3675,\, 0.6125\}$. 
We perform the same angle iterations as described above.  
Data with initial momenta $|\vec{P}_{\rm i}|/\mathrm{M}=\{0.245,0.490
\}$ are listed in both tables for completeness.

\begin{table}[htpb!]
\begin{tabular}{|c|c|c|c|c|}
\hline
Series & $|\vec{P}_{\rm i}|/\mathrm{M}$ & $\chi_{\rm i}$ & $\theta_{\rm M}$
       & $\theta_{\rm S}$ \\
\hline
Xp7P06 & $0.06125$   &  $0.7$  &  $(0.10000, 0.15100 )$   & $(0.15350, 0.17000)$ \\
Xp7P12 & $0.1225$   &  $0.7$  &  $(0.05000,0.07675)$   & $(0.07700, 0.08500)$ \\
Xp7P24 & $0.245$   &  $0.7$  &  $(0.04500, 0.04800)$   & $(0.04825, 0.05400)$  \\
Xp7P36 & $0.3675$   &  $0.7$  &  $(0.04000, 0.04225)$   & $(0.04250, 0.05300)$\\
Xp7P49 & $0.490$   &  $0.7$  & $(0.04000, 0.04200)$   & $(0.04250, 0.05400)$ \\
Xp7P61 & $0.6125$   &  $0.7$  &  $(0.03600, 0.04500)$ & $(0.04525, 0.04800)$ \\
\hline
\end{tabular}
\caption{\label{tab:RunsScatteringSpin0P7}
We summarize a set of simulations with different initial (linear) momenta $|\vec{P}_{\rm i}|/\mathrm{M}\in[0.06125,0.6125]$ for initial spin $\chi_{\rm i}=0.7$. We list the range of angles resulting in either merger $\theta_{\rm M}$, or scattering $\theta_{\rm S}$.}
\end{table}

The grid setup in the simulations is as follows. Each simulation is run on a three dimensional grid with outer boundary located at $x,y,z=\pm256\mathrm{M}$. To reduce computational cost, we leverage the symmetries of the binaries' setup and typically employ rotation symmetry and reflection symmetry in the z-direction. We use \textsc{Carpet} to employ box-in-box adaptive mesh refinement centered around the \bh{s}.
We set up seven refinement levels, where the innermost refinement levels are centered around each \bh{}.
The outermost refinement level has a resolution with step size $dx=1\mathrm{M}$.
Within consecutive refinement levels, we halve the step size such that the innermost refinement level has step size $dx=\frac{1}{64}\mathrm{M}$.
We set the Courant factor to \verb|dtfac| $=0.225$.
We set the \verb|time_refinement_factors| parameter, 
which controls how often 
each of the seven refinement levels is updated, 
to $[ 1, 1, 2, 4, 8, 16, 32]$.
With this setting, the rate at which the grid is updated is the same in the two outermost levels, and successively doubled thereafter in conjunction with doubling the spatial resolution on each refinement level.

In the simulations, we use two different setups for the refinement levels around the \bh{s}. The first (setup A) places the refinement boundaries at radii $r/\rm{M}=\{64.0, 16.0, 6.0, 3.0, 1.5, 0.75\}$ around the \bh{} centers. The second setup (setup B) places the refinement boundaries at radii $r/\rm{M}=\{64.0, 16.0, 4.0, 2.0, 1.0, 0.6\}$.
We use setup A in the majority of the simulations,
and setup B in some of the initially non-spinning Xp0P24 series.
The latter includes the zoom-whirl on which we perform a convergence test described in Sec.~\ref{sec:Setup_Convergence} and Appendix~\ref{sec:convergence_tests}.  

\subsection{Summary of convergence tests and error} \label{sec:Setup_Convergence}

To assess the numerical error of the simulation suites, we perform a convergence analysis on a set of three representative simulations. Namely, we run tests on one zoom-whirl simulation from the Xp0P24 series with initially non-spinning \bh{s} and two scattering simulations with high initial spin magnitude, $|\chi_{\rm i}|=0.7$, from the Xm7P49 and Xp7P49 series. The scattering simulations are selected such that one is run at the threshold angle, and the other is run at an angle much greater than the threshold angle. Zoom-whirls typically occur only for angles slightly below the threshold, and thus their behavior prior to merger can also be treated like a near threshold scattering simulation. With this set of convergence tests, we can assess the accuracy of simulations with different initial spins and incident angles. By using a zoom-whirl simulation with a lower initial momentum, we check that varying initial momenta and morphology do not notably impact convergence.

\begin{table}[htbp!]
    \centering
    \begin{tabular}{|c|c|c|c|c|}
    \hline
     Original Series & Morphology & $|\vec{P}_{\rm i}|/\mathrm{M}$ & $\theta$ & $\chi_{\rm i}$ \\
      \hline
     Xp0P24 & Zoom-Whirl & $0.245$ & $0.05800$ & 0.0 \\
     Xp7P49 & Scatter $\theta > \theta_{\rm th}$ & $0.490$ & $0.05685$ & $0.7$ \\
     Xm7P49 & Scatter $\theta=\theta_{\rm th}$ & $0.490$ & $0.05685$ & $-0.7$ \\
    \hline
    \end{tabular}
    \caption{\label{tab:Convergence_Summary} Parameters for suite of convergence tests. For each scenario we run a simulation with three different resolutions $dx_{\rm low}=1\mathrm{M}$,
    $dx_{\rm med}=0.95\mathrm{M}$, and $dx_{\rm high}=0.85\mathrm{M}$.}
\end{table}

The convergence tests are summarized in Table.~\ref{tab:Convergence_Summary}. For each case, we run three simulations with varying step size: (1) a low resolution simulation with step size $dx_{\rm low}=1\mathrm{M}$ (i.e. the standard value for simulations in this work), (2) a medium resolution simulation with step size $dx_{\rm med}=0.95\mathrm{M}$, and (3) a high resolution simulation with step size $dx_{\rm high}=0.85\mathrm{M}$. These step sizes refer to the outermost refinement level.
Near the \bh{s}, the simulations have respective step sizes
$dx_{\rm low}=\frac{1}{64}$M,
$dx_{\rm med}\approx\frac{1}{67}$M,
and
$dx_{\rm high}\approx\frac{1}{75}$M.

Using this set of simulations, we compute the relative error of the different reported observables, which are summarized in Table~\ref{tab:Error_Summary}. Errors for the Weyl scalar, $\Psi_{\mathrm{4,}22}$, are taken at the peak of the waveform. For the scattering tests, we report upper bounds on the error pre-encounter and post-encounter. For the zoom-whirl test, we report the same information for the first encounter along with upper bounds on errors for the remnant \bh{} after the merger. The pre-encounter values of the spin, $\chi$, and \bh{} angular momentum, $S$, are zero in the zoom-whirl test. Consequently, their percent error is poorly defined and, thus, listed as N/A. For a more detailed discussion and analysis of the convergence tests, error estimates, and uncertainty,
see Appendix~\ref{sec:convergence_tests}.

\begin{table}[htpb!]
\begin{tabular}{|c|c|c|c|c|c|c|c|c|c|}
\hline
 & \multicolumn{3}{c|}{Zoom-Whirl} & \multicolumn{2}{c|}{Scatter \footnotesize{$\theta > \theta_{\rm th}$} } & \multicolumn{2}{c|}{Scatter \footnotesize{$\theta=\theta_{\rm th}$} } \\
\hline
Data & Pre & Post & Merge &  Pre & Post & Pre & Post  \\
\hline
$\Psi_{4,22}$ & N/A & $0.14 \%$  & $1.6\%$ & N/A & $11.8 \%$ & N/A & $3.0\%$ \\
$m_{\rm irr}$ & $0.002 \%$ & $0.004 \%$  & $0.06 \%$ & $0.01 \%$ & $0.04 \%$  & $0.01\%$ & $0.6\%$ \\
$m$  & $0.001 \%$ & $0.005\%$   & $0.04 \% $ & $0.01\%$ & $0.04\%$ & $0.01\%$ & $0.3\%$ \\
$\chi$  & N/A & $5\%$  & $0.1 \%$ & $0.1\%$ & $0.4\%$ & $0.1\%$ & $6\%$ \\
$S$  & N/A & $5\%$ & $0.1\%$ & $0.1 \%$ & $0.5 \%$ & $0.1\%$ & $6\%$ \\
\hline
\end{tabular}
\caption{\label{tab:Error_Summary} Percent errors computed from the convergence tests. Errors for the gravitational radiation, $\Psi_{\mathrm{4,}22}$, are taken at the waveform peak. ``Pre'' refers to upper bounds on the error before scattering. ``Post'' refers to upper bounds on the error after scattering. ``Merger'' refers to upper bounds on the error of the remnant \bh{}. See Appendix~\ref{sec:convergence_tests} for further detail. }
\end{table}

\section{Results} \label{Sec:Results}
\subsection{Morphology of simulations} \label{Sec:Results_Morphology}

\begin{figure*}[htbp!]
    \begin{center}
    \includegraphics[width=2.\columnwidth]{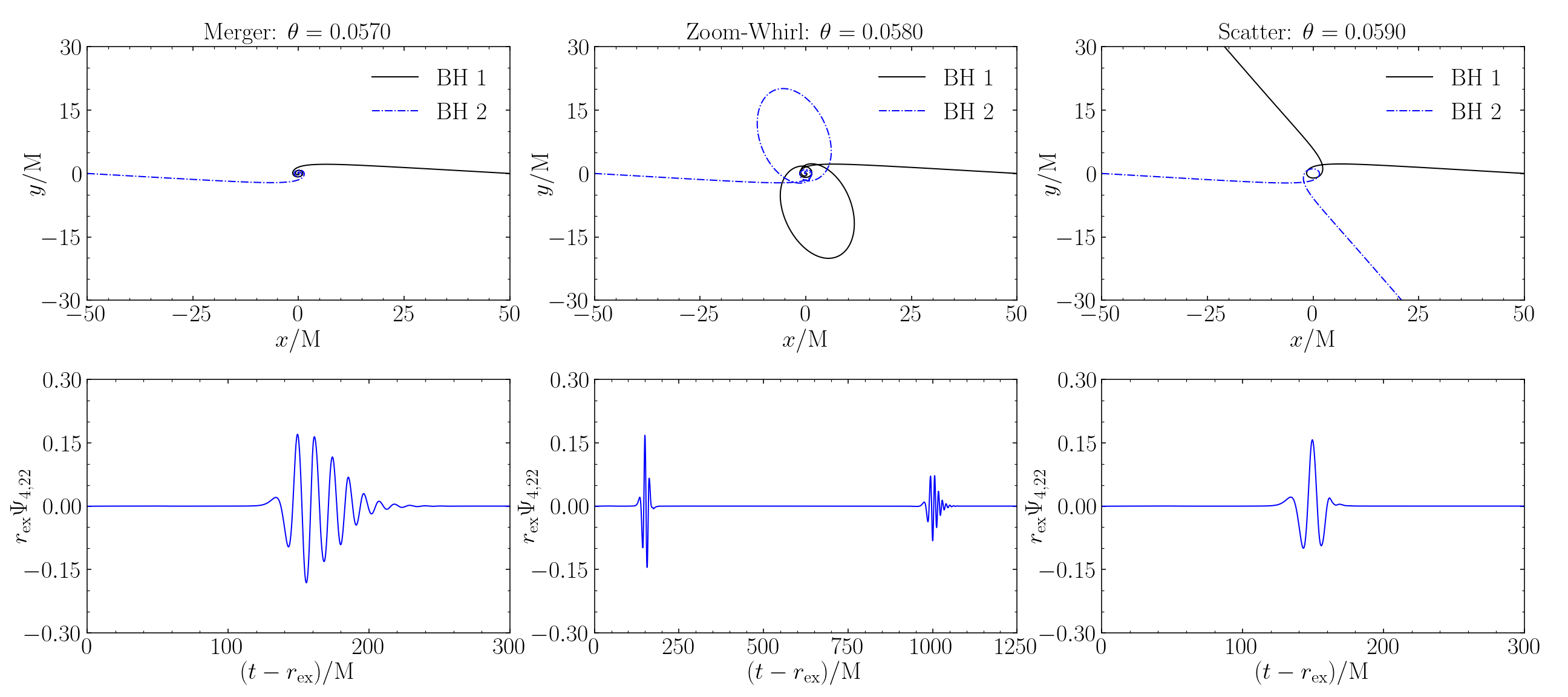}
    \caption{\label{fig:Boost245Spin00} Trajectory and gravitational waveform of \bh{} binaries with initial spin $\chi_{\rm i}=0.0$ and initial momentum $|\vec{P}_{\rm i}|=0.245\mathrm{M}$. From left to right, the panels depict systems with incident angles $\theta=0.0570$, $0.0580$, and $0.0590$ that result in a merger, zoom-whirl, and scattering, respectively. The threshold angle for this series of simulations is $\theta_{\rm th}=0.05825$. \underline{Top row:} Trajectory of the \bh{s} in the orbital (x-y) plane. \underline{Bottom row:} Gravitational radiation as given by the real part of the quadrupole of the Weyl scalar, $\Psi_{4,22}$, rescaled by the extraction radius, $\rex=100\mathrm{M}$. The time is shifted by the extraction radius.\\
    }
    \end{center}
\end{figure*}

\begin{figure*}[htbp!]
    \begin{center}
    \includegraphics[width=2.\columnwidth]{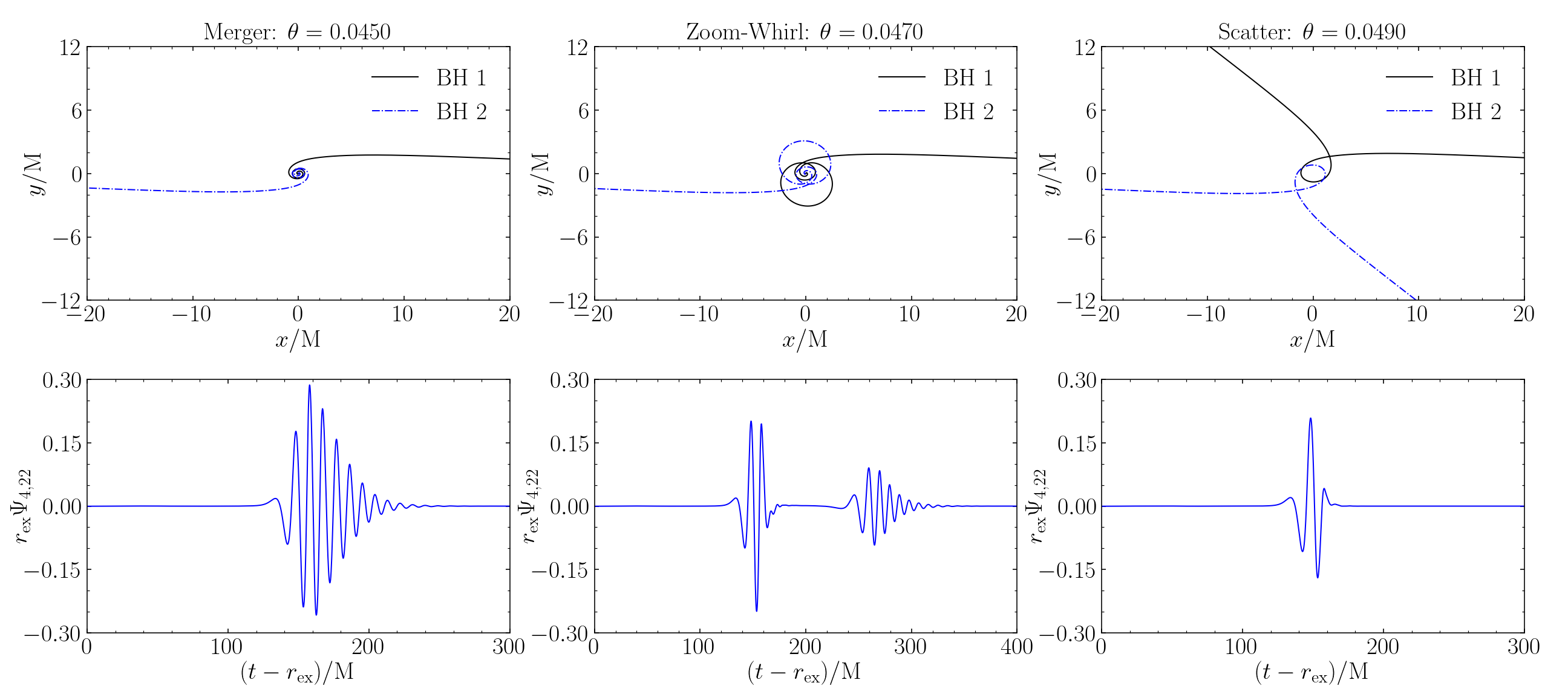}
    \caption{\label{fig:Boost245Spin07} Trajectory and gravitational waveform of \bh{} binaries with initial spin $\chi_{\rm i}=0.7$ and initial momentum $|\vec{P}_{\rm i}|=0.245\mathrm{M}$. From left to right the panels depict systems with incident angles $\theta=0.0450$, $0.0470$, and $0.0490$ that result in a merger, zoom-whirl, and scattering, respectively. The threshold angle for this series of simulations is $\theta_{\rm th}=0.04825$. \underline{Top row:} Trajectory of the \bh{s} in the orbital (x-y) plane. \underline{Bottom row:} Gravitational radiation as given by the real part of the quadrupole of the Weyl scalar, $\Psi_{4,22}$, rescaled by the extraction radius, $\rex=100\mathrm{M}$. The time is shifted by the extraction radius.\\
    }
    \end{center}
\end{figure*}

In this work, we study the behavior of binary \bh{s} that begin gravitationally unbound. The emission of energy in \gw{s} can cause the \bh{s} to become bound and, thus, undergo dynamical capture. Consequently, the binaries can exhibit three different morphologies depending on their incident angle: (1) mergers, in which the \bh{s} collide and form a remnant; (2) zoom-whirls, in which the \bh{s} undergo a series of small fast orbits (whirls) punctuated by larger eccentric orbits (zooms) until they eventually also merge; and (3) scattering (or hyperbolic orbits), in which the \bh{s} pass one another and escape to infinity. The \bh{s} merge at small angles, zoom-whirl (and merge) at intermediate angles, and scatter at large angles. We define the smallest angle for which the \bh{s} scatter to be the threshold angle, $\theta_{\rm th}$. We illustrate the three different morphologies in Figs.~\ref{fig:Boost245Spin00} and~\ref{fig:Boost245Spin07} by plotting the \bh{} trajectories and \gw{s} associated with examples of each case.

In Fig.~\ref{fig:Boost245Spin00}, we show examples of the three different morphologies for initially non-spinning, $\chi_{\rm i}=0.0$, \bh{s} with initial momentum $|\vec{P}_{\rm i}|=0.245\mathrm{M}$. From left to right, the panels depict the merger, zoom-whirl, and scattering of simulations with incident angles $\theta=0.0570$, $0.0580$, and $0.0590$, respectively. The threshold angle for this set of \bh{} parameters is $\theta_{\rm th}=0.05825$. In the top panels, we show the trajectories of the \bh{s} in the orbital (x-y) plane. In the bottom panels, we show the corresponding gravitational waveforms. Namely, we plot the real part of the quadrupole of the Weyl scalar, $\Psi_{\rm4,22}$, which quantifies outgoing gravitational radiation. We rescale the Weyl scalar by the extraction radius, $\rex=100\mathrm{M}$, to account for the radial fall off of the gravitational radiation. Furthermore, we shift the time by the extraction radius to account for the propagation delay of the radiation.

The merger waveform (left panel) follows the typical pattern of a \bh{} merger followed by an exponentially decaying ring-down.
The zoom-whirl waveform (middle panel) consists of two pieces. The first piece is a short pulse of radiation emitted during the whirl phase of the \bh{s}' encounter. In principle, there can be several pulses depending on the number of zoom-whirl cycles; however, in the simulation shown there is only one such cycle. The second piece corresponds to the zoom-whirl's merger and is qualitatively similar to the merger waveform discussed previously. The scattering waveform (right panel) shows a burst of radiation produced by the \bh{s}' close encounter.

In Fig.~\ref{fig:Boost245Spin07}, we display examples of the three different morphologies for \bh{s} with initial spin $\chi_{\rm i}=0.7$ and initial momentum $|\vec{P}_{\rm i}|=0.245\mathrm{M}$. 
From left to right, the panels depict the merger, zoom-whirl, and scattering of simulations with incident angles $\theta=0.0450$, $0.0470$, and $0.0490$, respectively. 
The threshold angle for this set of \bh{} parameters is $\theta_{\rm th}=0.04825$. In the top panels, we show the trajectories of the \bh{s} in the orbital (x-y) plane. In the bottom panels, we show the corresponding gravitational waveforms. 
We find qualitatively similar behavior to that of initially non-spinning \bh{s}.

\begin{figure}[htbp!]
    \begin{center}
    \includegraphics[width=1\columnwidth]{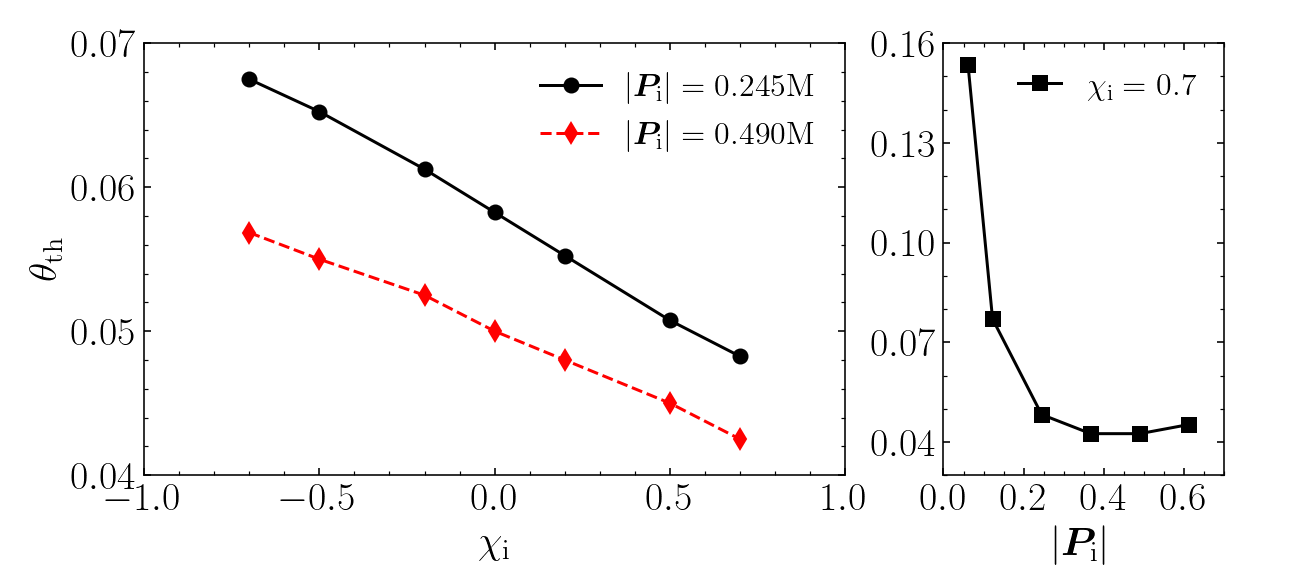}
    \caption{\label{fig:ThresholdAngle} Threshold angle, $\theta_{\rm th}$, as a function of the initial parameters.
    \underline{Left:}
    Dependence on the initial spin for initial momenta $|\vec{P}_{\rm i}|=0.245 \mathrm{M}$ and $|\vec{P}_{\rm i}|=0.490\mathrm{M}$.
    \underline{Right:}
    Dependence on the initial momentum for initial spin $\chi_{\rm i}=0.7$.}
    \end{center}
\end{figure}

\begin{figure}[htbp!]
    \begin{center}
    \includegraphics[width=1\columnwidth]{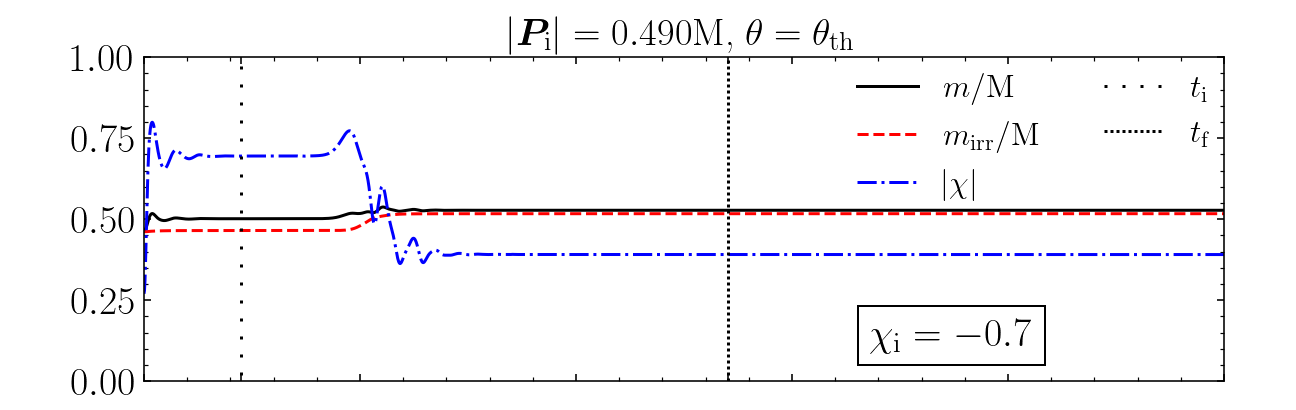}
    \includegraphics[width=1\columnwidth]{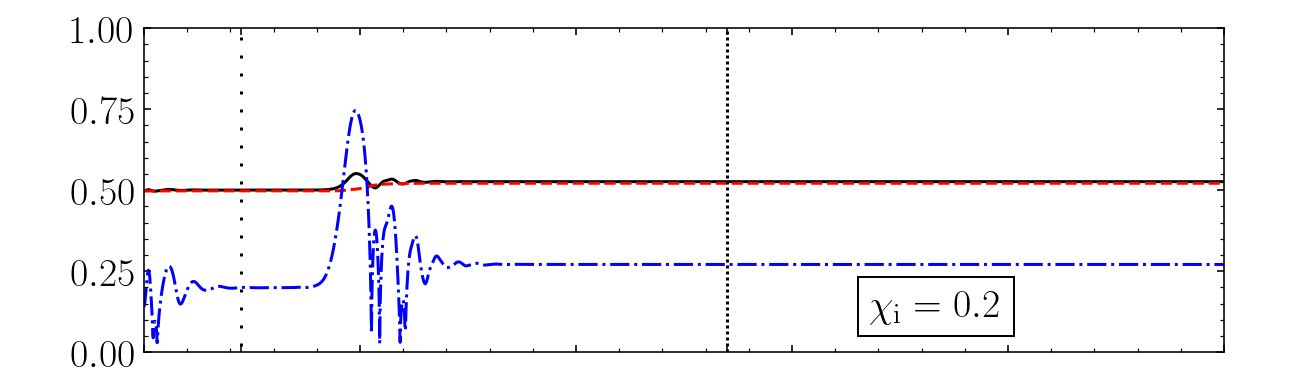}
    \includegraphics[width=1\columnwidth]{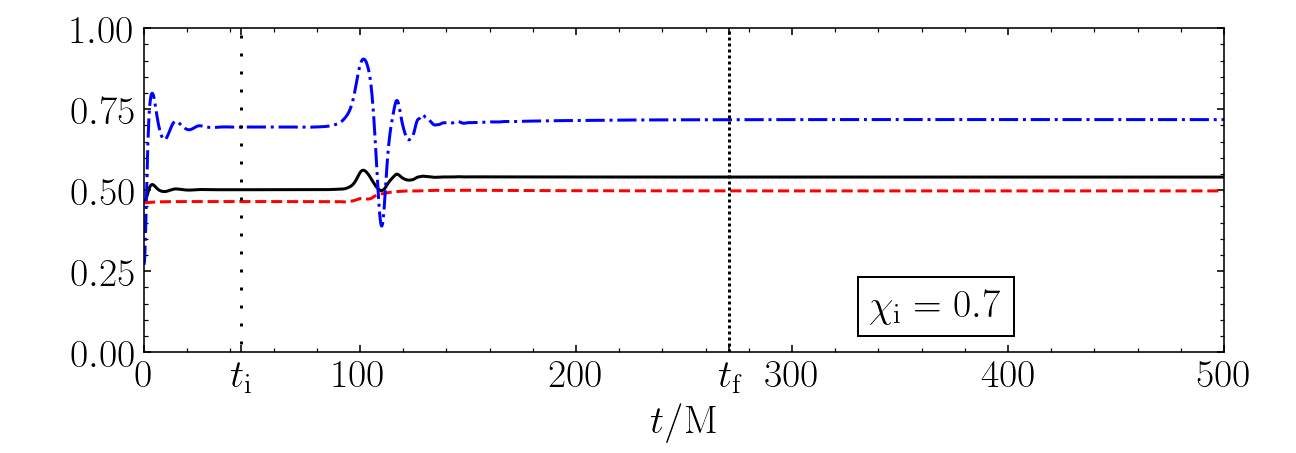}
    \caption{\label{fig:MassSpinvTime} Evolutions of the \bh{} mass, $m$, irreducible mass, $m_{\rm irr}$, and (dimensionless) spin magnitude, $|\chi|$, for \bh{s} scattering near the threshold angle, $\theta=\theta_{\rm th}$, with initial momentum $|\vec{P}_{\rm i}|=0.490 \mathrm{M}$. The dotted lines labeled $t_{\rm i}$ and $t_{\rm f}$ denote when initial and final quantities are measured. \underline{Top:} This plot ($\chi_{\rm i}=-0.7$) is typical of anti-aligned spins, where the spin's magnitude decreases causing the \bh{} mass and irreducible mass to approach in value. \underline{Middle:} This plot ($\chi_{\rm i}=0.2$) is typical of small aligned spins, where the spin increases, but makes negligible contribution to the \bh{} mass. \underline{Bottom:} This plot ($\chi_{\rm i}=0.7$) is typical of large aligned spins, where the spin change is marginal.}
    \end{center}
\end{figure}

Next, we analyze the dependence of the threshold angle on the initial spin and initial momentum; see Fig.~\ref{fig:ThresholdAngle}. In the left panel of Fig.~\ref{fig:ThresholdAngle}, we plot the threshold angle as a function of initial spin for initial momenta $|\vec{P}_{\rm i}|=0.245\mathrm{M}$ and $|\vec{P}_{\rm i}|=0.490\mathrm{M}$. For both initial momenta, the threshold angle decreases linearly as the initial spin increases. Moreover, for the same initial spin, the threshold angle decreases as the initial momentum increases. This trend of decreasing threshold angle with increasing initial momentum can also be seen in the right panel of Fig.~\ref{fig:ThresholdAngle}. Here, we plot the threshold angle as a function of the initial momentum for fixed initial spin $\chi_{\rm i}=0.7$. The behavior in this case appears linear for small initial momenta. However, for high initial momenta, the threshold angle appears to saturate or even increase slightly with initial momentum. 
We do not find a change in morphology in the convergence tests, even for systems near the threshold angle (see Appendix~\ref{sec:convergence_tests}). 
The error in the threshold angles is thus dominated by the spacing  between sampled angles (see Sec.~\ref{sec:Setup_Simulation_Suite}). 
For the remainder of this study, we focus on scattering \bh{s} and investigate how their spins and masses change due to their close encounter.

\subsection{Spin-up of scattered black holes}
\label{Sec:Results_Spinup}

One of the principal goals of this work is to study the change in spin experienced by scattering \bh{s} as a result of their encounter. The spin-up is studied by Refs.~\cite{Nelson:2019czq,Jaraba:2021ces} in initially non-spinning systems and by Refs.~\cite{Sperhake:2012me,Rodriguez-Monteverde:2024tnt} in initially spinning systems. We build upon this work by considering how different incident angles and initial momenta influence the change in spin in systems with initial spins $\chi_{\rm i}\in[-0.7,0.7]$.
We consider rotationally symmetric systems, where the spins are aligned or anti-aligned with the orbital angular momentum, i.e. along the z-axis.
By exploring a range of incident angles, we can compare systems scattered at the threshold angle, where we find that the change in spin is greatest.

\begin{figure*}[htpb!]
    \begin{center}

    \includegraphics[width=1\columnwidth]{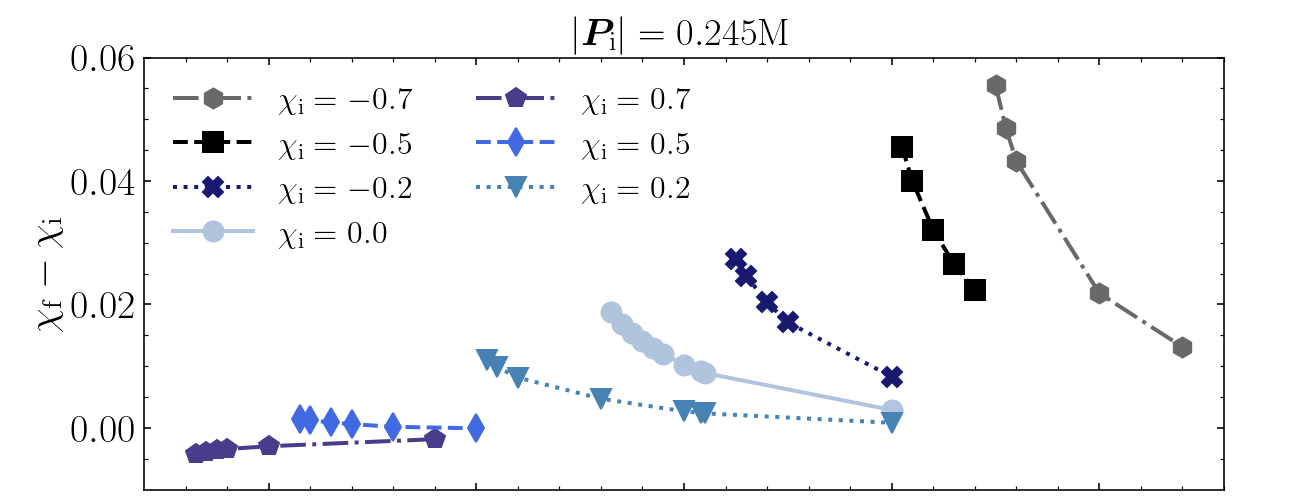}
    \includegraphics[width=1\columnwidth]{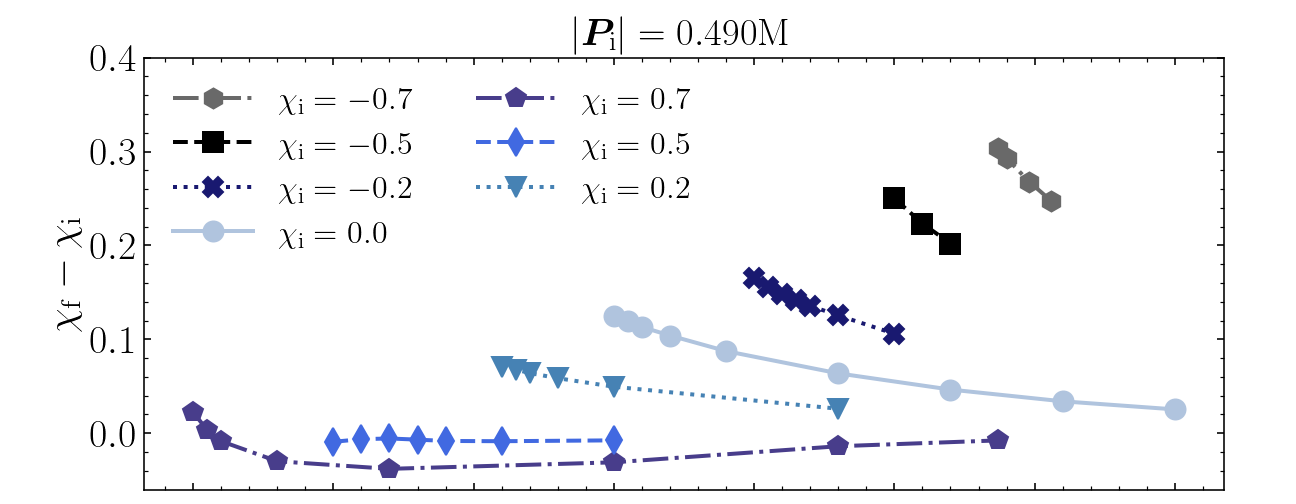}
    \includegraphics[width=1\columnwidth]{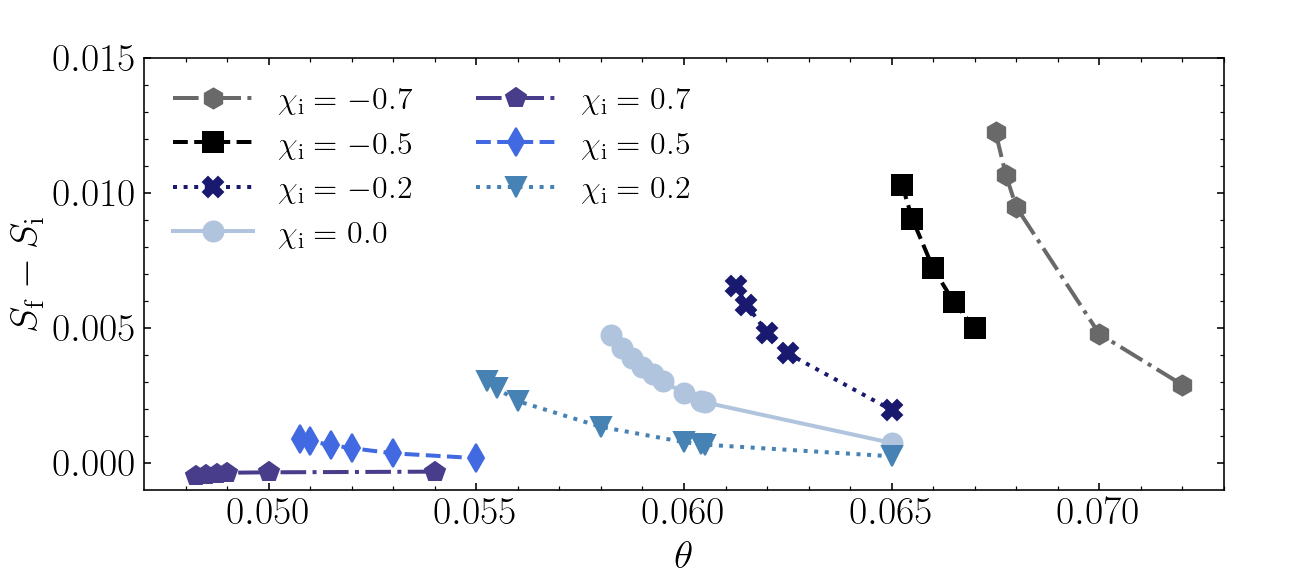}
    \includegraphics[width=1\columnwidth]{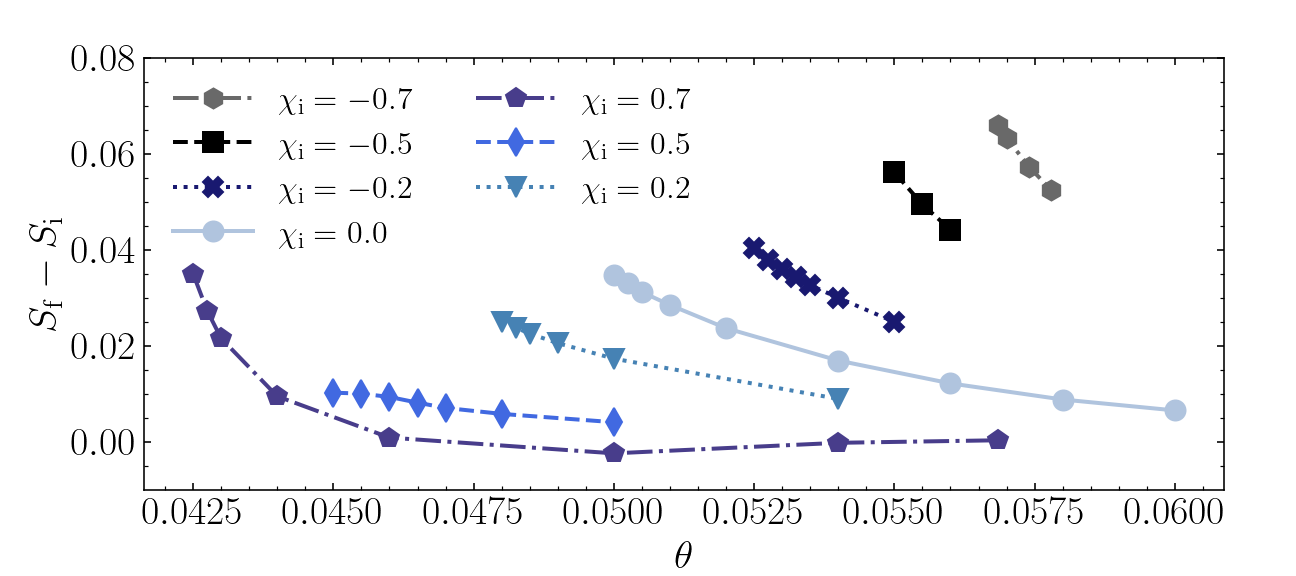}

    \caption{\label{fig:SpinvAngle} Change in the (dimensionless) spin (top panels) and \bh{} angular momentum (bottom panels) of scattering \bh{s} as a function of incident angle for initial momenta $|\vec{P}_{\rm i}|=0.245\mathrm{M}$ (left panels) and $|\vec{P}_{\rm i}|=0.490\mathrm{M}$ (right panels). We vary the \bh{s}' initial spin in the range $\chi_{\rm i}\in[-0.7,0.7]$ (as indicated).
    }
    \end{center}
\end{figure*}

In order to quantify the change in spin, $\chi_{\rm f}-\chi_{\rm i}$, we compute the initial spin, $\chi_{\rm i}$, and final spin, $\chi_{\rm f}$, of the scattering \bh{s} before and after an encounter according to Eq.~\eqref{eq:extract_spin}. We say that the \bh{s} spin-up if their change in spin is positive, $\chi_{\rm f}-\chi_{\rm i}>0$. For completeness, we also consider the change in the \bh{} angular momentum, $S_{\rm f}-S_{\rm i}$, where the initial \bh{} angular momentum, $S_{\rm i}$, and final \bh{} angular momentum, $S_{\rm f}$, are computed via Eq.~\eqref{eq:extract_angmom}.

In Fig.~\ref{fig:MassSpinvTime}, we plot representative evolutions of the spin magnitude as a function of time for three systems of scattering \bh{s}. The \bh{} mass and irreducible mass, which are also shown, are discussed in Sec.~\ref{Sec:Results_Mass}. Vertical dotted lines denote the initial time, $t_{\rm i}$, and final time, $t_{\rm f}$, at which we evaluate quantities prior to and following an encounter ($t\simeq100\mathrm{M}$). Each system has an incident angle equal to the threshold value and initial momentum $|\vec{P}_{\rm i}|=0.490\mathrm{M}$. The top, middle, and bottom panels depict systems with initial spins $\chi_{\rm i}=-0.7$, $0.2$, and $0.7$, respectively. In the top panel, the \bh{} spins up
as a result of the encounter, causing the spin magnitude to decrease. The magnitude decreases because the initial spin is negative (i.e. anti-aligned).
In the middle panel, the spin magnitude increases as the \bh{} spins up. However, the change is smaller than in the top panel. In the bottom panel, the change in the spin magnitude is negligible.

In Fig.~\ref{fig:MassSpinvTime}, we see that the spin magnitude oscillates around $t=0$M and $t\sim100$M.
The oscillation at the beginning of the simulation ($t=0$M) is due to gauge adjustments in the early evolution which can yield small modifications to the initial spin.
The oscillation around $t\sim100$M coincides with the \bh{s'} closest encounter during which they exert tidal forces on each other, and Eq.~\eqref{eq:extract_spin} may not be applicable.
We therefore evaluate the spin at a time $t_{\rm f}$ well after the encounter when the \bh{s} can be treated as isolated and Eq.~\eqref{eq:extract_spin} applies.

\begin{figure}[htbp!]
    \begin{center}
    \includegraphics[width=1\columnwidth]{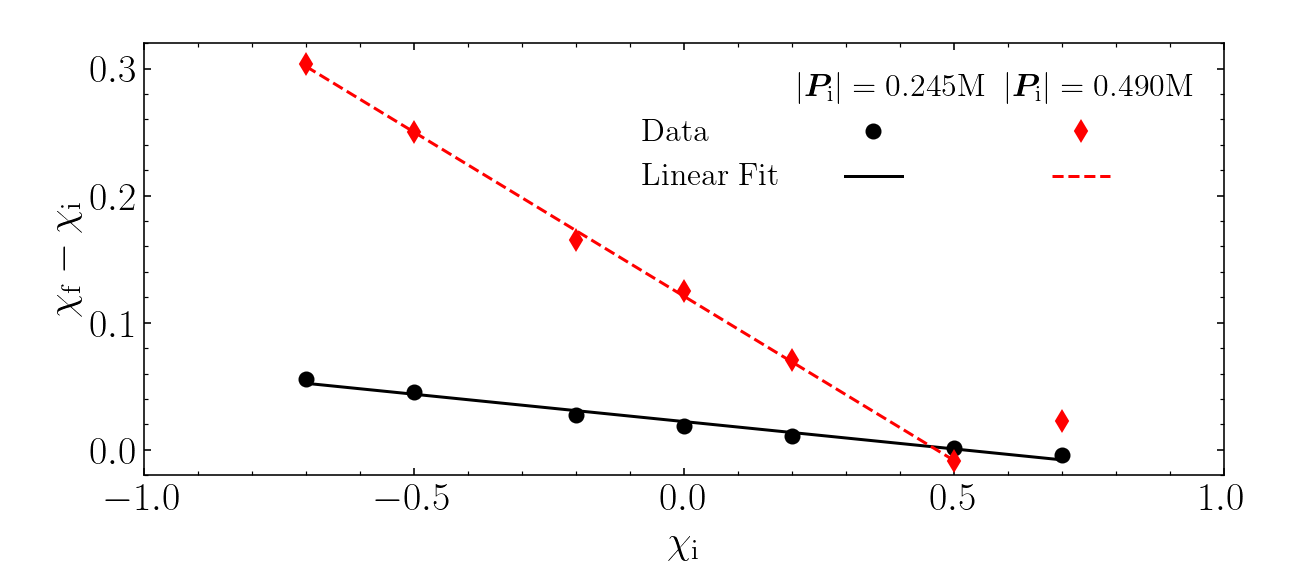}
    \caption{\label{fig:ThresholdAngle_Spin} Change in spin as a function of the initial spin of \bh{s} scattering near the threshold angle, $\theta=\theta_{\rm th}$. Data is shown for initial momenta $|\vec{P}_{\rm i}|=0.245\mathrm{M}$ and $|\vec{P}_{\rm i}|=0.490\mathrm{M}$. Points denote numerical data. Lines indicate linear best fits to the two data sets; see Eq.~\eqref{eq:best_fit_spin} and surrounding text for details.
    }
    \end{center}
\end{figure}

\begin{figure*}[htbp!]
    \begin{center}
    \includegraphics[width=2\columnwidth]{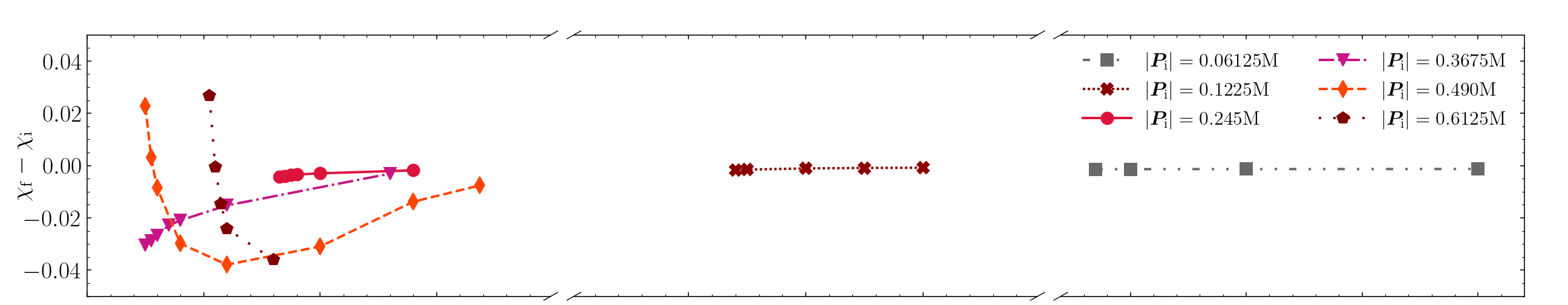}
    \includegraphics[width=2\columnwidth]{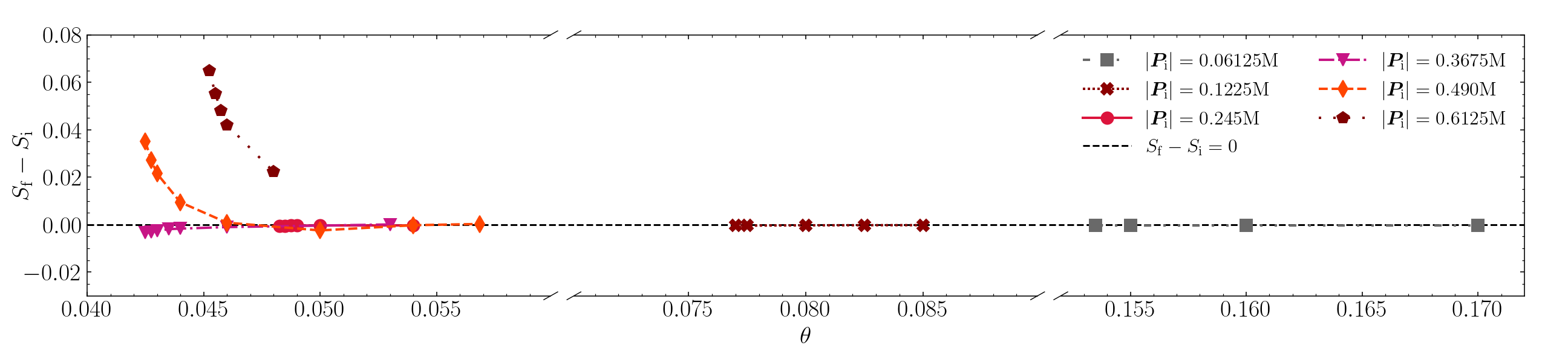}
    \caption{\label{fig:SpinvAngleBoosts} Change in the (dimensionless) spin (top panel) and \bh{} angular momentum (bottom panel) of scattering \bh{s} as a function of incident angle for fixed initial spin $\chi_{\rm i}=0.7$ and different initial momenta (as indicated). The dashed line corresponds to zero change in the \bh{} angular momentum, $S_{\rm f}-S_{\rm i}=0$.
    }
    \end{center}
\end{figure*}

When evaluating the initial and final values of the \bh{} parameters, we must be careful to avoid the above oscillations.
In principle, we could use the parameters listed in Tables~\ref{tab:RunsScatteringBoostP0245P0490} and~\ref{tab:RunsScatteringSpin0P7} for the initial spin. However, due to the initial gauge adjustments, these parameters can deviate from the initial spin found via Eq.~\eqref{eq:extract_spin} by approximately $\pm0.01$. Therefore, we recompute all quantities at an initial time ($t_{\rm i}=45\mathrm{M}$) placed about halfway between the start of the simulation and the encounter, when the spin is approximately constant. We compute final quantities at a time ($t_{\rm i}=270\mathrm{M}$) placed long enough after the encounter that the \bh{s} are isolated but not so late as to risk contamination from gravitational radiation reflected off the outer refinement boundary. In systems where the encounter occurs very early or late, we adjust the evaluation times to abide by these principals. We do not recompute the initial spin when $\chi_{\rm i}=0$. We estimate uncertainties in Appendix~\ref{sec:convergence_tests_Uncertainty} and find that they are typically smaller than the reported changes in spin and \bh{} angular momentum. Some exceptions exist for initial spins $\chi_{\rm i}\geq0.5$.

\subsubsection{Dependence on initial spin}

We first analyze the change in spin and \bh{} angular momentum for a set of systems with varying initial spin. In Fig.~\ref{fig:SpinvAngle}, we plot the change in spin (top panels) and \bh{} angular momentum (bottom panels) as a function of incident angle for different initial spins $\chi_{\rm i}\in[-0.7,0.7]$ and initial momenta $|\vec{P}_{\rm i}|/\mathrm{M}=\{0.245,0.490\}$. Each line corresponds to a series in Table~\ref{tab:RunsScatteringBoostP0245P0490}. The left panels show simulations with initial momentum $|\vec{P}_{\rm i}|=0.245\mathrm{M}$, and the right panels show simulations with initial momentum $|\vec{P}_{\rm i}|=0.490\mathrm{M}$.

For most initial spins, we find that the \bh{s} spin up,
and the increase in spin grows as the threshold angle is approached.
We note that
the spin magnitude increases in systems with aligned initial spins, while it decreases in systems with anti-aligned initial spins (i.e. the spins become less negative).
Moreover, the change in spin tends to decrease with increasing initial spin. Within the parameter range that we explore,
we find a maximum spin-up of about $\chi_{\rm f}-\chi_{\rm i}=0.3$.
This maximum is attained for an initial momentum $|\vec{P}_{\rm i}|=0.490\mathrm{M}$ and initial spin $\chi_{\rm i}=-0.7$.
The change in \bh{} angular momentum follows similar trends.

The only deviations from these trends occur for the positive initial spin of $\chi_{\rm i}=0.7$.
In particular, we find a negative change in spin
(i.e. a spin-down)
for some incident angles.
We also note that
the change in spin 
and \bh{} angular momentum 
is larger at the threshold angle than it is in systems with lower initial spin for initial momentum $|\vec{P}_{\rm i}|=0.490\mathrm{M}$.

In Fig.~\ref{fig:ThresholdAngle_Spin}, we plot the change in spin at the threshold angle as a function of initial spin for momenta $|\vec{P}_{\rm i}|=0.245\mathrm{M}$ and $|\vec{P}_{\rm i}|=0.490\mathrm{M}$. Heuristically, the data appears to follow a linear trend, which we model via least square fits,
\begin{equation} \label{eq:best_fit_spin}
    (\chi_{\rm f}-\chi_{\rm i})|_{\theta=\theta_{\rm th}}=a*\chi_{\rm i}+b \,,
\end{equation}
where $a$ and $b$ are fitting parameters. For momentum $|\vec{P}_{\rm i}|=0.245\mathrm{M}$, we find that $a=-0.043$ and $b=0.022$. The residual standard error is $0.0034$. For momentum $|\vec{P}_{\rm i}|=0.490\mathrm{M}$, we find that $a=-0.258$ and $b=0.121$. The residual standard error is $0.0042$. The data for initial spin $\chi_{\rm i}=0.7$ is not taken into consideration by the latter fit as
we find that it is within the estimated uncertainty (see Appendix~\ref{sec:convergence_tests_Uncertainty}).

The trends in Figs.~\ref{fig:SpinvAngle} and~\ref{fig:ThresholdAngle_Spin} remain consistent across both momenta. The primary distinction is that the changes in spin and \bh{} angular momentum are approximately an order of magnitude greater in systems with initial momentum $|\vec{P}_{\rm i}|=0.490\mathrm{M}$ than in those with $|\vec{P}_{\rm i}|=0.245\mathrm{M}$.

\subsubsection{Momentum dependence and spin-down}

Next, we analyze the dependence of the change in spin and \bh{} angular momentum on the initial momentum in systems with initial spin $\chi_{\rm i}=0.7$.
In Fig.~\ref{fig:SpinvAngleBoosts}, we plot the change in spin (top panel) and \bh{} angular momenta (bottom panel) as a function of incident angle
for initial momenta $|\vec{P}_{\rm i}|/\mathrm{M}\in[0.06125,0.6125]$.
Each line corresponds to a series in Table~\ref{tab:RunsScatteringSpin0P7}. The black dashed line denotes zero change in the \bh{} angular momentum.

We first consider the change in spin.
For small initial momenta $|\vec{P}_{\rm i}|\leq0.245\mathrm{M}$, we find that the change in spin is consistent with zero.
For intermediate initial momentum,
$|\vec{P}_{\rm i}|=0.3675$,
the change in spin becomes negative close to the threshold.
For initial momenta $|\vec{P}_{\rm i}|\geq0.490\mathrm{M}$, the change in spin is
positive near the threshold,
it is negative at intermediate scattering angles,
and it approaches zero far from the threshold.

When considering these trends, we must be cognizant of numerical error. At large incident angles, we find an uncertainty of $\Delta (\chi_{\rm f}-\chi_{\rm i})\simeq 0.003$ for initial spin $\chi_{\rm i}=0.7$ (see Appendix~\ref{sec:convergence_tests_Uncertainty}). This uncertainty is smaller than the spin-down that we observe at large and intermediate incident angles for initial momenta $|\vec{P}_{\rm i}|\geq0.3675\mathrm{M}$. We can thus be confident that spin-down is a physical phenomena. However, near the threshold angle, we find an uncertainty of $\Delta (\chi_{\rm f}-\chi_{\rm i})\simeq0.04$ for initial spin $\chi_{\rm i}=0.7$. This uncertainty is larger than the spin-up that we observe for small angles in Fig.~\ref{fig:SpinvAngleBoosts}.
While seemingly large, this spin-up is thus consistent with zero within numerical error.
However, given that this behavior matches what we find with smaller initial spins at small incident angles, it is likely qualitatively correct, even
if not quantitatively.

\begin{figure}[htbp!]
    \begin{center}
    \includegraphics[width=1\columnwidth]{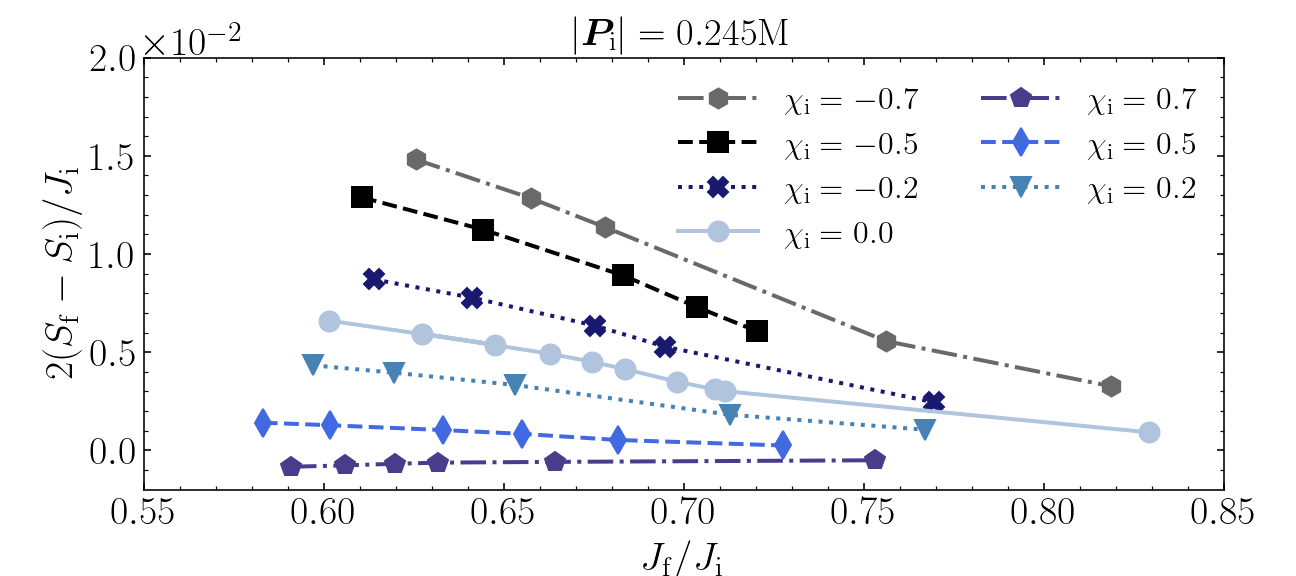}
    \includegraphics[width=1\columnwidth]{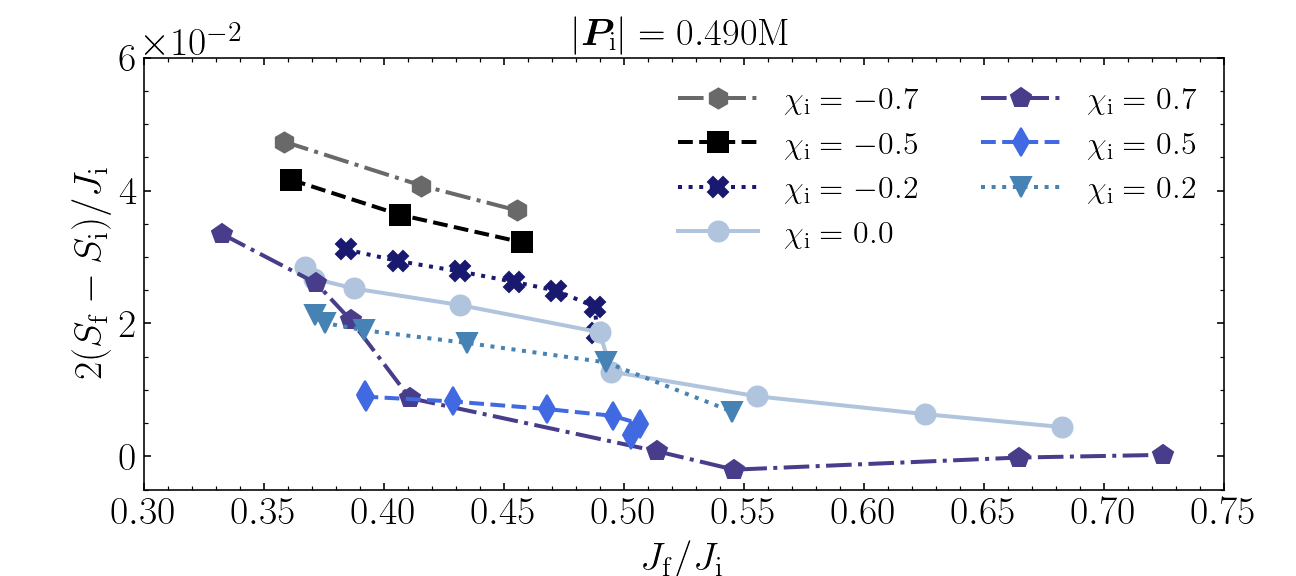}
    \includegraphics[width=1\columnwidth]{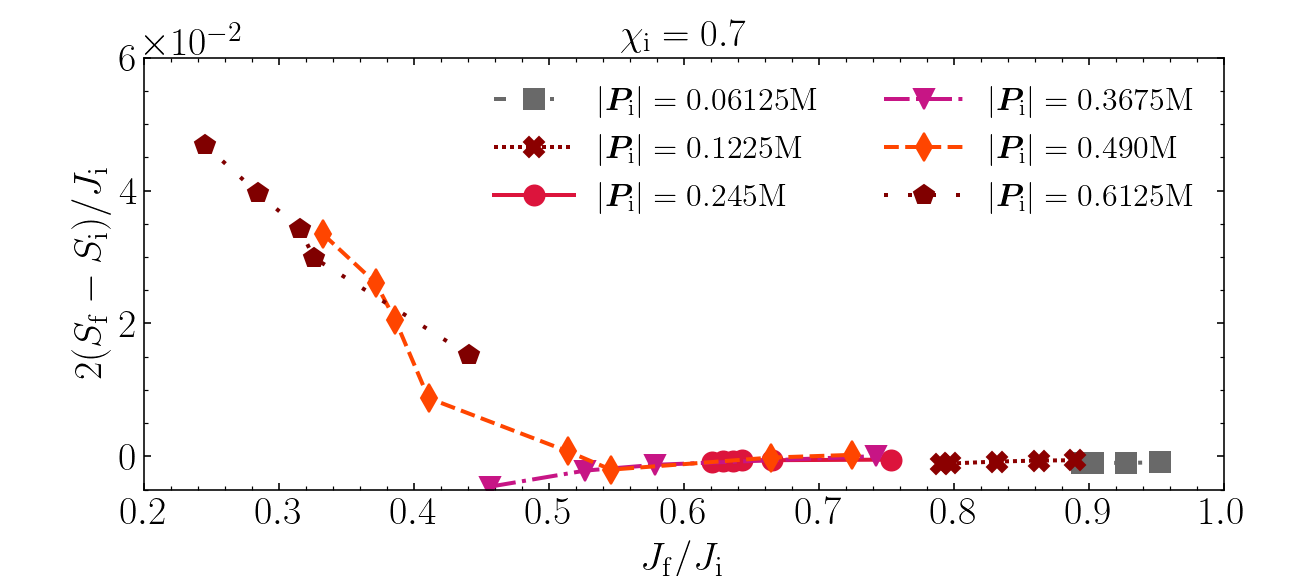}
    \caption{\label{fig:Spinup_Efficiency} Spin-up efficiency, Eq.~\eqref{eq:efficiency}, of scattering \bh{s}
    as a function of the    
    ratio between the final and initial orbital angular momentum, $J_{\rm f}/J_{\rm i}$. \underline{Top:} Results for initial momentum $|\vec{P}_{\rm i}|=0.245\mathrm{M}$. \underline{Middle:} Results for initial momentum $|\vec{P}_{\rm i}|=0.490\mathrm{M}$. \underline{Bottom:} Results for different initial momenta with initial spin $\chi_{\rm i}=0.7$.}
    \end{center}
\end{figure}

We now consider the change in \bh{} angular momentum (see bottom of Fig.~\ref{fig:SpinvAngleBoosts}). The change in \bh{} angular momentum is qualitatively similar to the change in spin in that there is little change for initial momenta $|\vec{P}_{\rm i}|\leq0.3675\mathrm{M}$, but for initial momenta $|\vec{P}_{\rm i}|\geq0.490\mathrm{M}$, the final \bh{} angular momentum increases near the threshold angle. However, unlike the change in spin, the change in the \bh{} angular momentum is always positive or consistent with zero. Some points still appear slightly negative, but these changes are smaller than the estimated uncertainties. For an initial spin of $\chi_{\rm i}=0.7$, we find an uncertainty of $\Delta (S_{\rm f}-S_{\rm i})\simeq0.01\mathrm{M}^2$ near the threshold angle and an uncertainty of $\Delta (S_{\rm f}-S_{\rm i})\simeq0.001\mathrm{M}^2$ far from the threshold angle (see Appendix~\ref{sec:convergence_tests_Uncertainty}). Furthermore, we find that the increase in \bh{} angular momentum observed at small angles increases with increasing initial momentum.

The observation that the \bh{} angular momentum never decreases hints at the origin of spin-down. A \bh{'s} spin is equal to its angular momentum divided by its mass squared; see Eq.~\eqref{eq:extract_angmom}. Since the \bh{} angular momentum never decreases, the spin down we find must be attributable to an increase in the \bh{} mass. We further discuss the behavior of the \bh{} mass in Sec.~\ref{Sec:Results_Mass}.

\subsubsection{Spin-up efficiency}

The change in the \bh{} angular momentum originates from a decrease in the system's orbital angular momentum, $J$,
which is radiated in
\gw{s} and partially re-absorbed by the \bh{s}.
Following Refs.~\cite{Nelson:2019czq,Rodriguez-Monteverde:2024tnt}, we seek to understand this process by computing the spin-up efficiency,
\begin{equation} \label{eq:efficiency}
  2(S_{\rm f}-S_{\rm i})/J_{\rm i} \, ,
\end{equation}
that quantifies the
fraction of the initial orbital angular momentum, $J_{\rm i}$, transferred into the \bh{s}' angular momenta.
The initial
and the final orbital angular momenta are given in Eqs.~\eqref{eq:Orbital_AngMom_initial} and~\eqref{eq:Orbital_AngMom_final},
respectively.

In Fig.~\ref{fig:Spinup_Efficiency}, we plot the spin-up efficiency against the ratio of the final to the initial orbital angular momentum, $J_{\rm f}/J_{\rm i}$. The top and middle panels show the results for different initial spins $\chi_{\rm i}\in[-0.7,0.7]$ with initial momenta $|\vec{P}_{\rm i}|=0.245\mathrm{M}$ and $|\vec{P}_{\rm i}|=0.490\mathrm{M}$, respectively. Each line corresponds to a series in Table~\ref{tab:RunsScatteringBoostP0245P0490}. The bottom panel shows the results for initial spin $\chi_{\rm i}=0.7$ and varying initial momenta $|\vec{P}_{\rm i}|/\mathrm{M}\in[0.06125,0.6125]$. Each line corresponds to a series in Table~\ref{tab:RunsScatteringSpin0P7}.

In the top and middle panels of Fig.~\ref{fig:Spinup_Efficiency}, we find that the spin-up efficiency typically increases with decreasing (or more negative) initial spin. In the bottom panel of Fig.~\ref{fig:Spinup_Efficiency}, we find that the spin-up efficiency increases with increasing initial momentum. We also find that the fraction of the orbital angular momentum retained by the system decreases with increasing initial momentum.

\begin{figure*}[htbp!]
    \begin{center}
    \includegraphics[width=1\columnwidth]{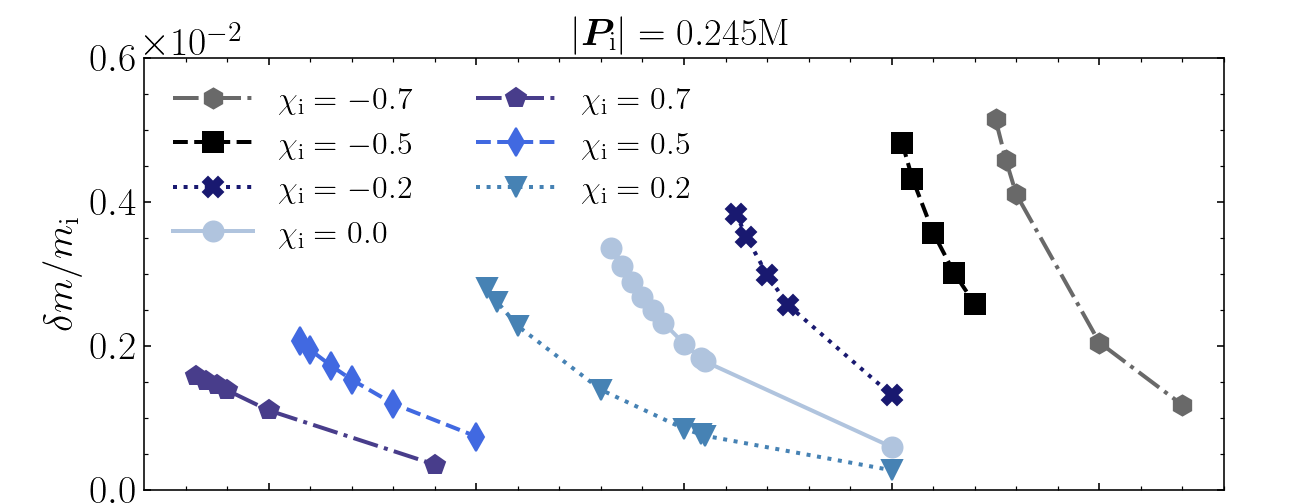}
    \includegraphics[width=1\columnwidth]{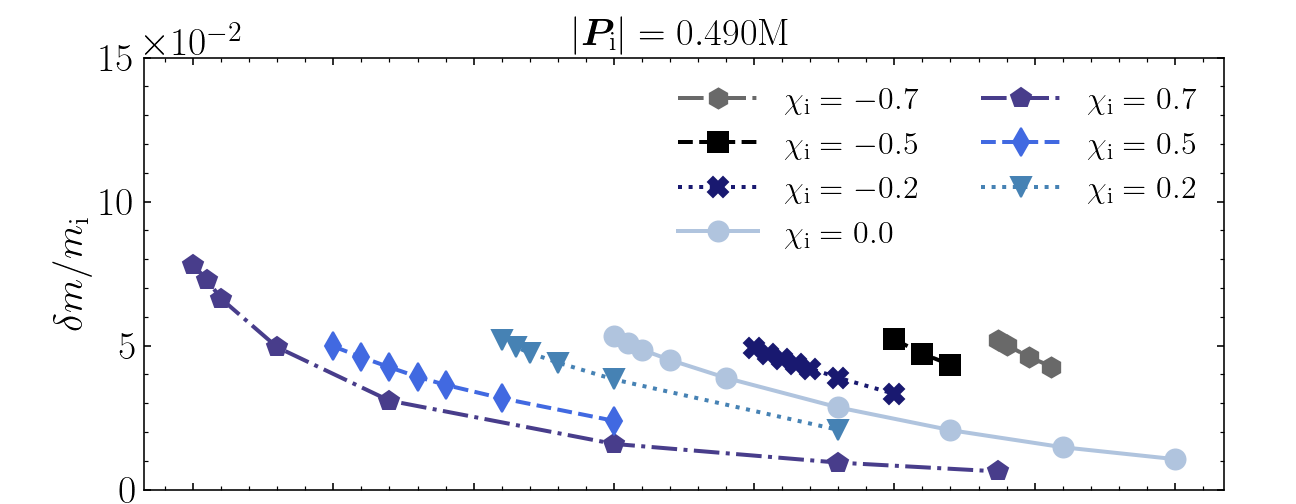}
    \includegraphics[width=1\columnwidth]{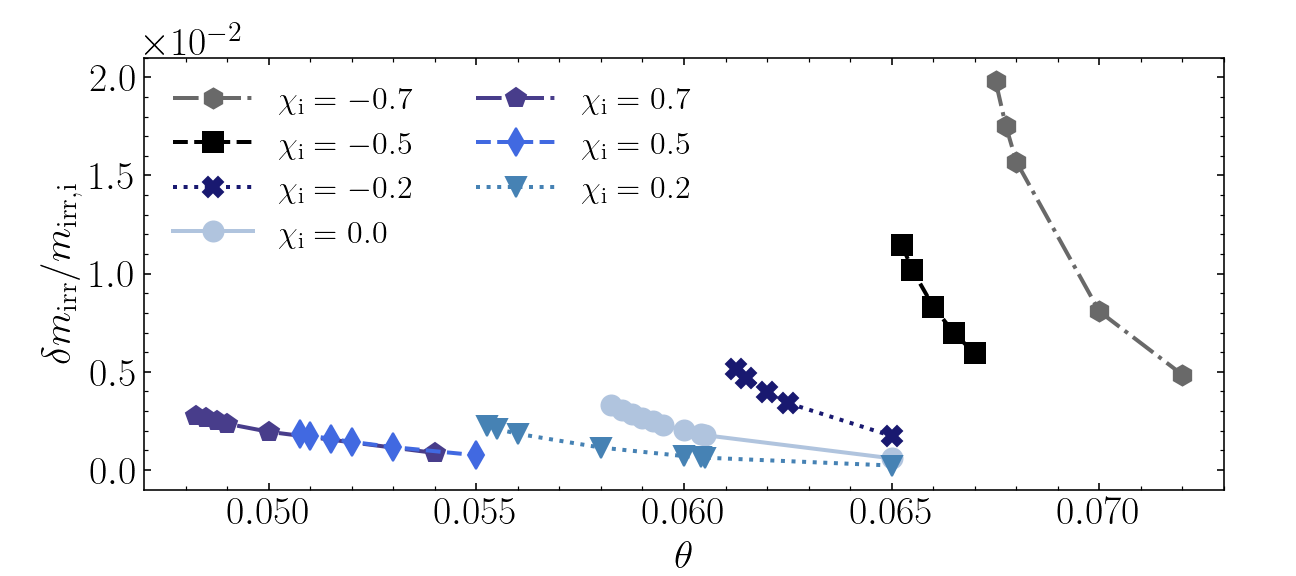}
    \includegraphics[width=1\columnwidth]{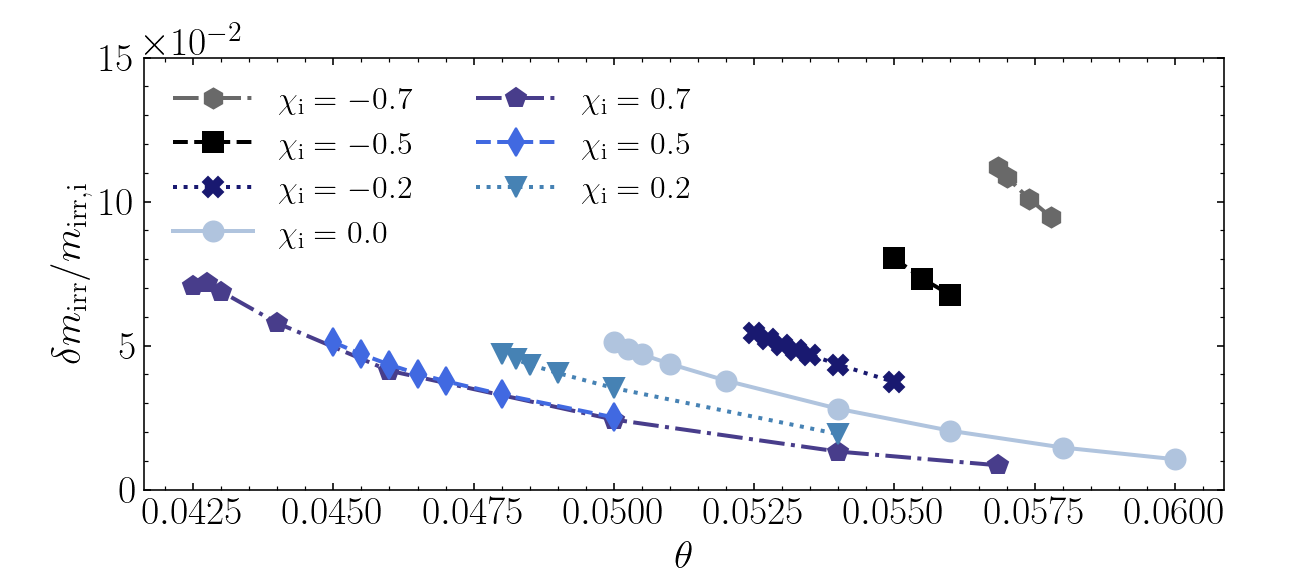}
    \caption{\label{fig:MassvAngle} Relative change in the \bh{} mass (top panels) and irreducible mass (bottom panels) of scattering \bh{s} as a function of incident angle for initial momenta $|\vec{P}_{\rm i}|=0.245\mathrm{M}$ (left panels) and $|\vec{P}_{\rm i}|=0.490\mathrm{M}$ (right panels). We vary the \bh{s}' initial spin in the range $\chi_{\rm i}\in[-0.7,0.7]$ (as indicated).}
    \end{center}
\end{figure*}

Throughout Fig.~\ref{fig:Spinup_Efficiency}, we can see that the spin-up efficiency tends to be larger in systems that retain a smaller fraction of the initial orbital angular momentum. Across all panels, we find a maximum spin-up efficiency of just under $5\%$. This is attained in systems with initial momentum $|\vec{P}_{\rm i}|=0.490\mathrm{M}$ and initial spin $\chi_{\rm i}=-0.7$ (see middle panel), as well as in systems with initial momentum $|\vec{P}_{\rm i}|=0.6125\mathrm{M}$ and initial spin $\chi_{\rm i}=0.7$ (see bottom panel). Note that while some efficiencies appear slightly negative, the corresponding changes in \bh{} angular momentum are consistent with zero within numerical error.

\subsection{Mass-gain} \label{Sec:Results_Mass}

In addition to the spin and \bh{} angular momentum, we find that scattering also leads to a change in the \bh{} mass and irreducible mass; see Fig.~\ref{fig:MassSpinvTime}.
The mass-gain of scattering \bh{s} has been studied
with the effective-one-body approach~\cite{Chiaramello:2024unv} and in simulations of ultra-relativistic \bh{} scattering~\cite{Sperhake:2012me}.
Here, we conduct a numerical analysis of the mass-gain in scattering \bh{s} across a wide parameter space. Furthermore, we comment on how the mass-gain relates to the observed spin behavior detailed in Sec.~\ref{Sec:Results_Spinup}.

\begin{figure}[htbp!]
    \begin{center}
    \includegraphics[width=1\columnwidth]{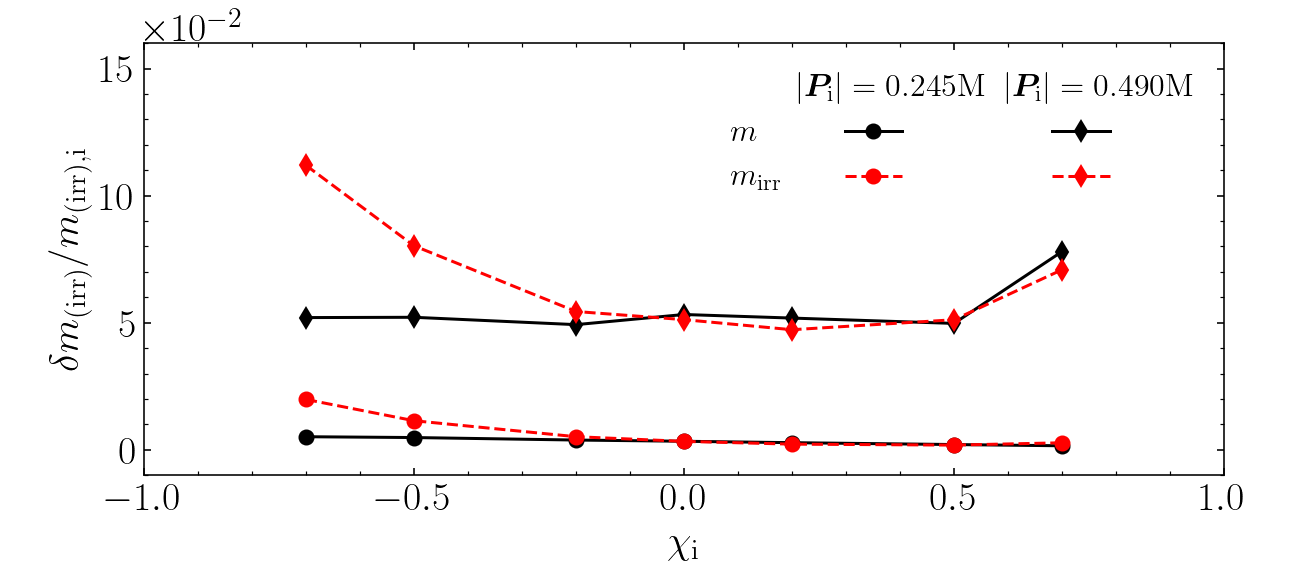}
    \caption{\label{fig:ThresholdAngle_Masses_Spin} Relative change in the \bh{} mass (solid lines) and irreducible mass (dashed lines) as a function of the initial spin of \bh{s} scattering near the threshold angle, $\theta=\theta_{\rm th}$. Data is shown for initial momenta $|\vec{P}_{\rm i}|=0.245\mathrm{M}$ and $|\vec{P}_{\rm i}|=0.490\mathrm{M}$.}
    \end{center}
\end{figure}

As discussed in Sec.~\ref{sec:Setup_Observables}, the \bh{} mass is composed of the irreducible mass and the \bh{} angular momentum (or spin) according to Eq.~\eqref{eq:mchris_angmom}. Here we consider the behavior of both the \bh{} mass and the irreducible mass. The \bh{} mass is computed via Eq.~\eqref{eq:extract}, and the irreducible mass is computed directly by the \verb|AHFinderDirect| thorn. Since the \bh{} mass and irreducible mass are positive definite, we report their evolution as relative changes,
\begin{equation} \label{eq:mass_relative_change}
     \frac{\delta m_{\rm (irr)}}{m_{\rm (irr), i}}=\frac{m_{\rm (irr),f}-m_{\rm (irr), i}}{m_{\rm (irr), i}} \,,
\end{equation}
where $m_{\rm (irr)}$ refers either to the \bh{} mass or the irreducible mass. 
$m_{\rm (irr),i}$ and $m_{\rm(irr),f}$ refer to their initial and final values before and after scattering,
evaluated at the times $t_{\rm i}$ and $t_{\rm f}$ indicated in Fig.~\ref{fig:MassSpinvTime}.
We estimate uncertainties in Appendix~\ref{sec:convergence_tests_Uncertainty} and find that they are typically smaller than the relative changes in \bh{} mass and irreducible mass reported,
except for small initial momenta and positive initial spins.

\begin{figure*}[htbp!]
    \begin{center}
    \includegraphics[width=2\columnwidth]{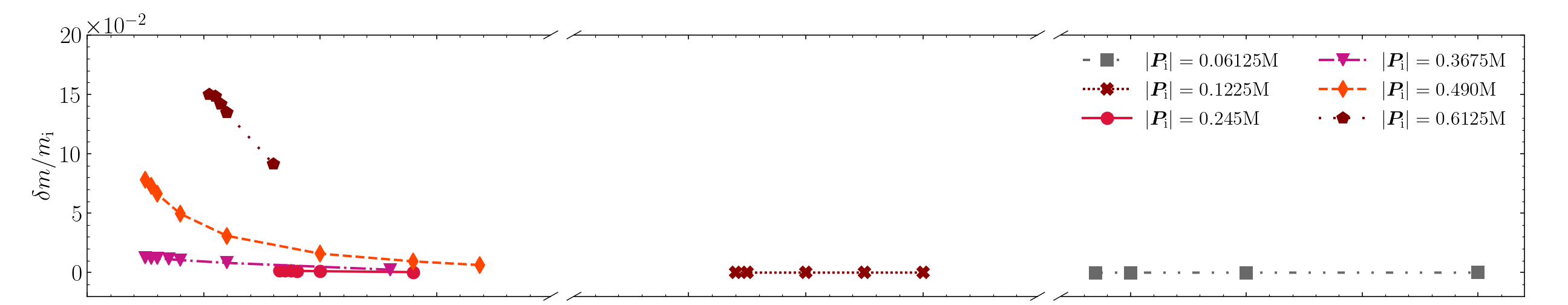}
    \includegraphics[width=2\columnwidth]{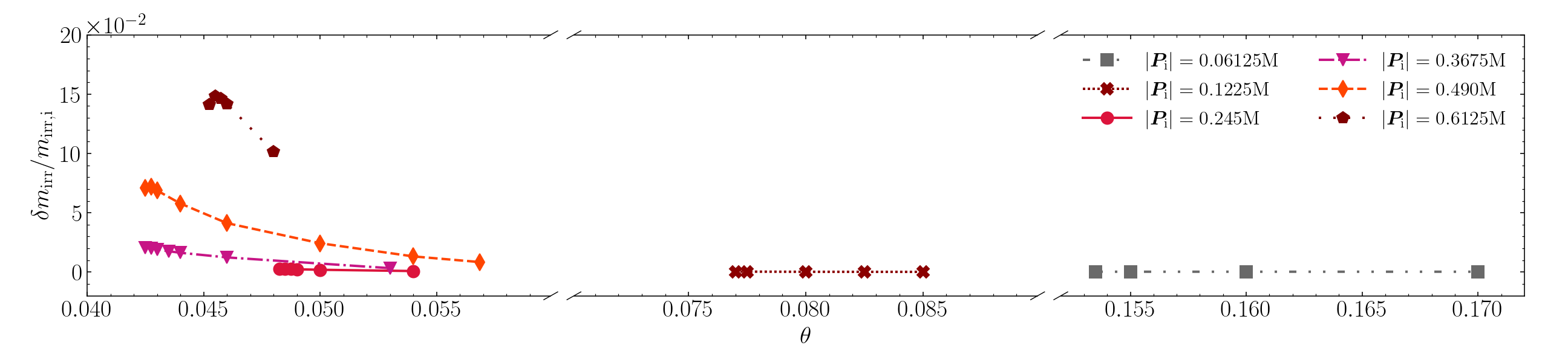}
    \caption{\label{fig:MassvAngleBoosts} Relative change in the \bh{} mass (top panel) and irreducible mass (bottom panel) of scattering \bh{s} as a function of incident angle for fixed initial spin $\chi_{\rm i}=0.7$ and different initial momenta (as indicated).}
    \end{center}
\end{figure*}

\subsubsection{Initial spin dependence}

We first analyze the relative change in the \bh{} mass and irreducible mass for a set of systems with varying initial spin. In Fig.~\ref{fig:MassvAngle}, we plot the relative change in the \bh{} mass (top panels) and irreducible mass (bottom panels) as a function of incident angle for different initial spins $\chi_{\rm i}\in[-0.7,0.7]$ and initial momenta $|\vec{P}_{\rm i}|/\mathrm{M}=\{0.245,0.490\}$. Each line corresponds to a series from Table~\ref{tab:RunsScatteringBoostP0245P0490}. The left panels show simulations with initial momentum $|\vec{P}_{\rm i}|={0.245\mathrm{M}}$, and the right panels show simulations with initial momentum $|\vec{P}_{\rm i}|=0.490\mathrm{M}$.

We find that the trends in the \bh{} mass and irreducible mass are similar to those found for the change in spin.
Namely, they always increase
and this gain in mass becomes
larger as the threshold angle is approached.
Furthermore, the increase is usually larger for smaller (i.e. more negative) initial spins, and it is roughly an order of magnitude greater in systems with initial momentum $|\vec{P}_{\rm i}|=0.490\mathrm{M}$ than in those with  
$|\vec{P}_{\rm i}|=0.245\mathrm{M}$.
The largest change in the \bh{} mass, for this simulation series, is about $8\%$.
It is obtained
for an initial momentum $|\vec{P}_{\rm i}|=0.490\mathrm{M}$ and initial spin $\chi_{\rm i}=0.7$.
The largest change in the irreducible mass is about $11\%$. It is found for an initial momentum $|\vec{P}_{\rm i}|=0.490\mathrm{M}$ and initial spin $\chi_{\rm i}=-0.7$.

In Fig.~\ref{fig:ThresholdAngle_Masses_Spin}, we plot the relative change in the \bh{} mass and irreducible mass at the threshold angle as a function of initial spin for initial momenta $|\vec{P}_{\rm i}|=0.245\mathrm{M}$ and $|\vec{P}_{\rm i}|=0.490\mathrm{M}$.
This plot highlights several unique trends.
The change in the \bh{} mass for
both initial momenta
is approximately uniform across different initial spins,
with the exception of $\chi_{\rm i}=0.7$, which is larger. 
For both initial momenta, the change in \bh{} mass is approximately equal to the change in irreducible mass for positive initial spins. However, simulations with negative initial spin have greater changes in their irreducible mass than in their \bh{} mass.

Although most clearly visible in Fig.~\ref{fig:ThresholdAngle_Masses_Spin}, close inspection of Fig.~\ref{fig:MassvAngle} reveals that simulations with negative initial spin have a greater relative change in irreducible mass than in \bh{} mass across the incident angles considered.
Physically, this behavior is a consequence of \bh{} thermodynamics and
the decrease in spin magnitude which occurs in systems with negative initial spin (see Sec.~\ref{Sec:Results_Spinup}).
In all simulations, the irreducible mass increases by some amount because the second law of \bh{} thermodynamics forbids the horizon area and, ergo, the irreducible mass from decreasing \cite{Hawking:1971tu,Bekenstein:1972tm}. Conversely, in simulations where
the initial spin is negative, the spin-up leads to a decrease in spin magnitude.
Therefore, while the irreducible mass term in Eq.~\eqref{eq:mchris_angmom} increases due to an encounter,
the contribution from the spin becomes smaller.
Consequently the relative change in the \bh{} mass can be
 smaller than that of the irreducible mass.

We can see examples of this behavior in Fig.~\ref{fig:MassSpinvTime}, which shows representative time evolutions of the \bh{} mass, irreducible mass, and spin magnitude for scattering \bh{s} with a variety of initial spins (see Sec.~\ref{Sec:Results_Spinup}). In the top panel, where the initial spin is negative, but the initial spin magnitude is large ($|\chi_{\rm i}|=0.7$), there is initially a clear gap between the \bh{} mass and irreducible mass. However, due to the spin-up, the spin magnitude decreases after the encounter, and thus the gap between the \bh{} mass and irreducible mass decreases. In the middle panel, where the initial spin is positive and small, there is little gap between the \bh{} mass and irreducible mass either before or after the encounter. In the bottom panel, where the initial spin is positive and large, there is a noticeable gap between the \bh{} mass and irreducible mass both before and after the encounter.

\begin{figure}[htbp!]
    \begin{center}
    \includegraphics[width=1\columnwidth]{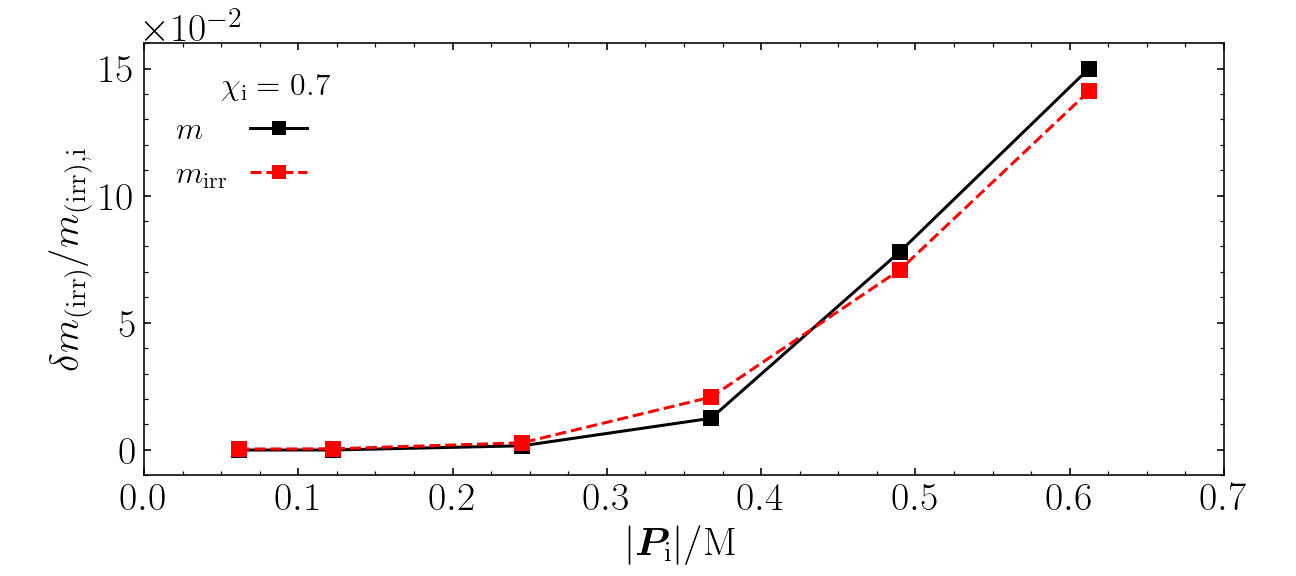}
    \caption{\label{fig:ThresholdAngle_Masses_Momentum} Relative change in the \bh{} mass (solid line) and irreducible mass (dashed line) as a function of the initial momentum of \bh{s} scattering near the threshold angle, $\theta=\theta_{\rm th}$. The \bh{s} have initial spin $\chi_{\rm i}=0.7$.}
    \end{center}
\end{figure}

\subsubsection{Momentum dependence}

Next, we further analyze the dependence of the mass-gain on the initial momentum in systems with initial spin $\chi_{\rm i}=0.7$. In Fig.~\ref{fig:MassvAngleBoosts}, we plot the relative change in the \bh{} mass (top panel) and irreducible mass (bottom panel) as a function of incident angle for initial spin $\chi_{\rm i}=0.7$ and several initial momenta $|\vec{P}_{\rm i}|/\mathrm{M}\in[0.06125,0.6125]$. Each line corresponds to a series from Table~\ref{tab:RunsScatteringSpin0P7}.

We find that the relative changes in the \bh{} mass and irreducible mass display similar behavior. For initial momenta $|\vec{P}_{\rm i}|\leq0.245\mathrm{M}$, there is little change in either quantity, but for initial momenta $|\vec{P}_{\rm i}|\geq0.3675\mathrm{M}$, both quantities increase in a manner that becomes larger as the threshold angle is approached. Furthermore, these increases become larger with increasing initial momentum.

In Fig~\ref{fig:ThresholdAngle_Masses_Momentum}, we plot the relative change in the \bh{} mass and irreducible mass at the threshold angle as a function of initial momentum. Here we see again that the relative change in both quantities increases with the increasing initial momentum. The smallest initial momenta produce changes close to zero, and the largest initial momenta yield changes up to about $15\%$.

\section{Conclusions and Outlook} \label{Sec:Conclusion}

In this work, we have investigated the mass-gain and spin-up (or spin-down) in the scattering of equal-mass, spinning \bh{s} and determined their dependence on the initial spins and initial (linear) momenta.
Therefore, we have performed a series of simulations
in which we considered \bh{s} with initial spin in the range
$\chi_{\rm i}\in [ -0.7,0.7]$
for initial momenta $|\vec{P}_{\rm i}|/\mathrm{M}\in\{0.245,0.490\}$.
Astrophysically, they are perhaps the most interesting choices as most \bh{s} observed with ground-based gravitational-wave detectors merge into \bh{s} with spins around $\chi\sim0.7$ \cite{LIGOScientific:2025slb}.
For the highest initial spin in our simulations, $\chi_{\rm i}=0.7$, we have varied the \bh{s'} initial momenta in the range
$|\vec{P}_{\rm i}|/\mathrm{M}\in[0.06125,0.6125]$.

For each set of parameters, we vary the incident angle and identify the threshold between scattering and merger.
We observe that the threshold angle decreases with increasing initial spin and initial momentum.
It appears to saturate for high initial momenta.

We have found a change in the spin of the scattered \bh{s}, as compared to their initial spin, due to the transfer of orbital angular momentum.
In particular, we have observed both a spin-up and a spin-down depending on the initial conditions.
We have found that the spin-up is largest for angles near the threshold value,
large initial momenta
and negative initial spins
(i.e. anti-aligned with the orbital angular momentum).
The change in spin, evaluated at the threshold, decreases {\textit{linearly}} with increasing initial spin.
Across the simulations, we identified
a maximum spin-up of
$\chi_{\rm f}-\chi_{\rm i}=0.3$
for an initial spin $\chi_{\rm i}=-0.7$
and initial momentum $|\vec{P}_{\rm i}|=0.490\mathrm{M}$.
Furthermore, we found a decrease of the (dimensionless) spin
(or ``spin-down'')
in simulations with moderate to high
positive initial spin $\chi_{\rm i}=0.7$.
Note, however, that this spin-down is a consequence of an increased \bh{} mass rather than a decrease of the (dimensionful)
\bh{} angular momentum. 

The change in the \bh{} angular momentum,
as compared to its initial value before the \bh{s'} scattering,
exhibits trends that are similar to those of the (dimensionless) spin.
However, unlike the change in spin, we find that the change in \bh{} angular momentum is always positive
because it accounts for the increase in the \bh{} mass.

The spin-up efficiency
increases with decreasing (or more negative) initial spin and with increasing initial momentum.
Across the simulations, we find a maximum spin-up efficiency of just under $5\%$ in systems with
initial momentum $|\vec{P}_{\rm i}|=0.490\mathrm{M}$ and negative initial spin $\chi_{\rm i}=-0.7$,
as well as in systems with initial momentum $|\vec{P}_{\rm i}|=0.6125\mathrm{M}$ and initial spin $\chi_{\rm i}=0.7$.

In all simulations, we have observed an increase in the irreducible mass and in the \bh{} mass after scattering.
The gain in mass is largest for scattering angles near the threshold,
large initial momenta,
and negative initial spins.
Across the simulations, we find a maximum increase of about $15\%$ in both the \bh{} mass and irreducible mass for initial momentum $|\vec{P}_{\rm i}|=0.6125\mathrm{M}$ and initial spin $\chi_{\rm i}=0.7$.
For binaries with positive initial spins
(i.e., aligned with the orbital angular momentum),
the changes in the \bh{} mass and the irreducible mass are comparable.
In contrast, in binaries with negative initial spins
(i.e., anti-aligned)
the change in the irreducible mass is larger than that of the \bh{} mass.
This difference in behavior occurs because the spin magnitude decreases as a result of scattering in systems with negative initial spin.
Due to the decline in spin magnitude, the increase in \bh{} mass is less than the increase in irreducible mass; see Eq.~\eqref{eq:mchris_angmom}.

Looking ahead, much can be done to further explore these phenomena.
It would be interesting to investigate the spin-up (or spin-down) and mass-gain
in unequal-mass binaries of initially spinning \bh{s},
or in precessing binaries with unequal spins.
Given that we find the most interesting behavior for negative initial spins, it would be instructive to further explore their evolution in a wider range of initial momenta.
For example,
early work on this topic has shown that
the threshold angle becomes insensitive to the initial spin in the scattering of ultra-relativistic \bh{s}~\cite{Sperhake:2012me},
and it would be interesting to understand how this limit is approached.

While it is possible to fine-tune large changes in the masses and spins of scattering \bh{s}, the impact of these changes on the dynamics of dense clusters is less clear.
Successive encounters might produce cumulative effects.
However, the spin magnitude can either increase or decrease depending on the initial spin alignment.
For example, Ref.~\cite{Chiaramello:2024unv} 
found that \bh{s} scattering in dense clusters
typically experience a decrease in their spin magnitude if initially $|\chi_{\rm i}|>0.2$.
It would be interesting to see if the same behavior is found when different methods and initial conditions are used.

With the advance of fully relativistic N-body simulations in numerical relativity~\cite{Lousto:2007rj,Campanelli:2007ea,Imbrogno:2021xrh,Ficarra:2023zjc,Ficarra:2024jen,Heinze:2025usf,Bamber:2025gxj},
it could also be interesting to isolate hyperbolic encounters as they dynamically arise within a cluster and look for these effects.
With improved detectors and modeling for hyperbolic encounters \cite{Garcia-Bellido:2021jlq,Kerachian:2023gsa, Kocsis:2006hq,Mukherjee:2020hnm,Morras:2021atg,Bini:2023gaj}, scattering \bh{s} in dense clusters may someday be detectable via \gw{s}.

Finally, many extensions to general relativity include dynamical scalar fields coupled to curvature invariants.
Their excitation or amplitude can have a strong qualitative dependence on the spin; see e.g. Refs.~\cite{Dima:2020yac,Herdeiro:2020wei,Doneva:2023oww,Richards:2025ows,Elley:2022ept}.
There may be interesting phenomena to be explored resulting from the change in spin due to scattering in theories beyond general relativity.

\section*{Acknowledgments} \label{Sec:Acknowledgements}
We thank
C.~-~H.~Cheng,
D.~Ferguson,
R.~Haas,
and
H.~O.~da~Silva,
for insightful discussions and comments.
The authors acknowledge support provided by the National Science Foundation under NSF Award No.~OAC-2004879,
No.~OAC-2411068 and No.~PHY-2409726.
This material is based upon work supported by the National Science Foundation Graduate Research Fellowship Program under Grant No. DGE-2146755, awarded to H.~K. Any opinions, findings, and conclusions or recommendations expressed in this material are those of the author(s) and do not necessarily reflect the views of the National Science Foundation.
We acknowledge the Texas Advanced Computing Center (TACC) at the University of Texas at Austin for providing HPC resources on Frontera via allocation PHY22041.
This research used resources provided by the Delta research computing project, which is supported by the NSF Award No. OAC-2005572
and the State of Illinois.
This research was supported in part by the Illinois Computes project which is supported by the University of Illinois Urbana-Champaign and the University of Illinois System.
This work used the open-source softwares
\textsc{xTensor}~\cite{xAct:web,Brizuela:2008ra},
the \ETK~\cite{maxwell_rizzo_2025_15520463,Loffler:2011ay, Zilhao:2013hia}, and
\canuda~\cite{witek_2023_7791842}.

\appendix

\section{Convergence Tests and Error Estimates}~\label{sec:convergence_tests}

We conduct convergence tests to assess the numerical error of the simulations presented in this study.
Therefore, we run representative simulations from the Xp0P24, Xp7P49, and Xm7P49 series in Table~\ref{tab:RunsScatteringBoostP0245P0490} with step sizes $dx_{\rm low}> dx_{\rm med}>dx_{\rm high}$.
In the outermost refinement levels,
the step sizes are $dx_{\rm low}=1.0\mathrm{M}$, $dx_{\rm med}=0.95\mathrm{M}$, and $dx_{\rm high}=0.85\mathrm{M}$.
Within the inner refinement levels, the step sizes are successively halved.
The simulations presented in the main text use the step size, $dx_{\rm low}$.

As we increase the resolution of a simulation, the quantities it computes should
converge towards the ``true'' solution.
The rate at which they approach the solution
is related to the simulations' order of convergence, $n$,
which is used to compute the convergence factor,
\begin{equation}
    Q_{n}(dx_{\rm low},dx_{\rm med},dx_{\rm high})=\frac{dx_{\rm low}^n - dx_{\rm med}^n}{dx_{\rm med}^n - dx_{\rm high}^n} \, .
\end{equation}
In the simulations, we use fourth order finite differencing for spatial derivatives and a fourth order Runge-Kutta scheme for stepping forward in time.
At refinement boundaries we
use a second order interpolation in time and fifth order in space.
Therefore, we may find a mixed convergence order in the simulations.
The convergence factors for fourth, third and second order convergence are, respectively,
$Q_{4}(1.0,0.95,0.85)=0.634$,
$Q_{3}=0.586$ and $Q_{2}=0.542$.
To verify that a quantity converges at the expected rate, we plot the difference between its values at low and medium resolution, and the difference between its values at medium and high resolution multiplied by the convergence factor.
In the following, we use the notation $q_{\rm low}-q_{\rm med}$ and $Q_{n}(q_{\rm med}-q_{\rm high})$ to refer to these differences in quantities and ``$q$'' to refer to the quantity calculated at a given step size.

We compute the relative error (or ``percent'' error) for different quantities by using the highest resolution simulation as a reference,
\begin{equation} \label{eq:Error}
    \%\mathrm{Error}=100 \, \left| \frac{q_{\rm low}-q_{\rm high}}{q_{\rm high}} \right|
\,.
\end{equation}
Note that Eq.~\eqref{eq:Error} explicitly uses units of $\%$.
Meaningful evaluation of the percent error requires care when handling quantities that are not positive definite, as the percent error will diverge if the higher resolution value
goes to zero.
This phenomenon is especially problematic when considering gravitational waveforms, where the sign changes frequently.
To address this issue, the error estimates for the Weyl scalar are calculated at the peak value of a given waveform within the relevant time interval.

In the Xp0P24 test, the spin is initially zero,
so we only report errors for the spin and \bh{} angular momentum after the encounter, once the \bh{s} have spun up. For other quantities, where this issue is less pervasive, we plot the percent error as a function of time and then report the maximum value attained within a given region.

In this study, we explore a broad set of parameters and phenomenology including mergers, zooms-whirls, and scattering.  
It is important to understand how the numerical accuracy varies across the different morphologies and parameters.
Therefore, we conduct tests on systems that occupy extremities in the parameter space: (1) a zoom-whirl with initial spin $\chi_{\rm i}=0$, initial momentum $|\vec{P}_{\rm i}|=0.245\mathrm{M}$, and incident angle slightly below the threshold value, $\theta \lesssim \theta_{\rm th}$, from the Xp0P24 series;
(2) a scattering with initial spin $\chi_{\rm i}=0.7$, initial momentum $|\vec{P}_{\rm i}|=0.490\mathrm{M}$, and large incident angle, $\theta>\theta_{\rm th}$, from the Xp7P49 series;
and (3) another scattering with initial spin $\chi_{\rm i}=-0.7$, initial momentum $|\vec{P}_{\rm i}|=0.490\mathrm{M}$, and incident angle equal to the threshold angle, $\theta=\theta_{\rm th}$, from the Xm7P49 series.

In the following, we show convergence and error plots for the quadrupole mode of the Weyl scalar, the irreducible mass, the \bh{} mass, the spin, and the \bh{} angular momentum. Summaries of the suite of convergence tests and computed errors are given in Tables~\ref{tab:Convergence_Summary} and~\ref{tab:Error_Summary}.

\subsection{Zoom-Whirl} \label{sec:convergence_tests_ZW}

We first present the convergence test for the zoom-whirl simulation from the Xp0P24 series with incident angle $\theta=0.0580$, initial spin $\chi_{\rm i}=0$, and initial momentum $|\vec{P}_{\rm i}|=0.245\mathrm{M}$.
This zoom-whirl undergoes one close encounter prior to the merger, during which the initially non-spinning \bh{s} acquire a spin of $\chi\sim0.2$.
This first encounter and the merger are separated by a ``zoom'' trajectory, where the \bh{s} are comparatively widely separated (see Sec.~\ref{Sec:Results_Morphology}).
Depending on the resolution, the separation of the \bh{s} during the zoom phase of the encounter varies.
While the trajectories realign during the final inspiral, this variation in the \bh{s'} separation shifts the time of merger between the simulations.
For this reason, we perform the convergence test in two blocks.
The first block focuses on the first encounter, whereas the second focuses on the merger and the remnant \bh{}.
In the second block, the times are shifted by $t_{\rm S}=t_{\mathrm{ merge,}\, dx}-t_{\mathrm{ merge, \,} dx_{\rm low}}$, to align the data at the time of merger in the low resolution run, $dx=dx_{\rm low}=1.0$.

\begin{figure}[htbp!]
    \begin{center}
    \includegraphics[width=1\columnwidth]{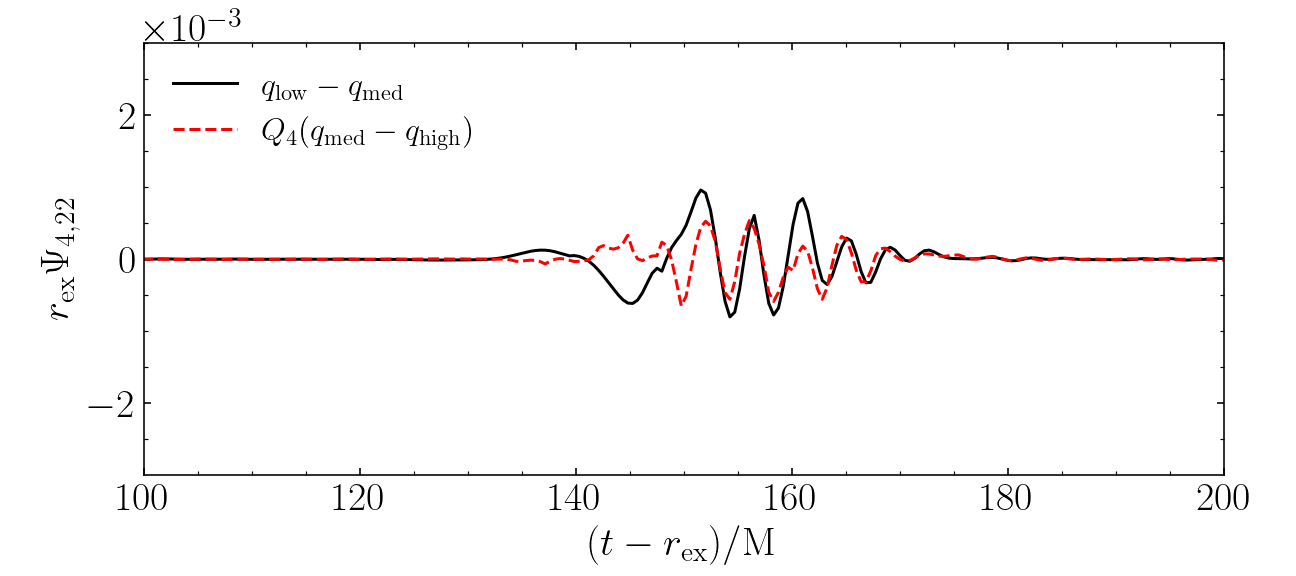}
    \includegraphics[width=1\columnwidth]{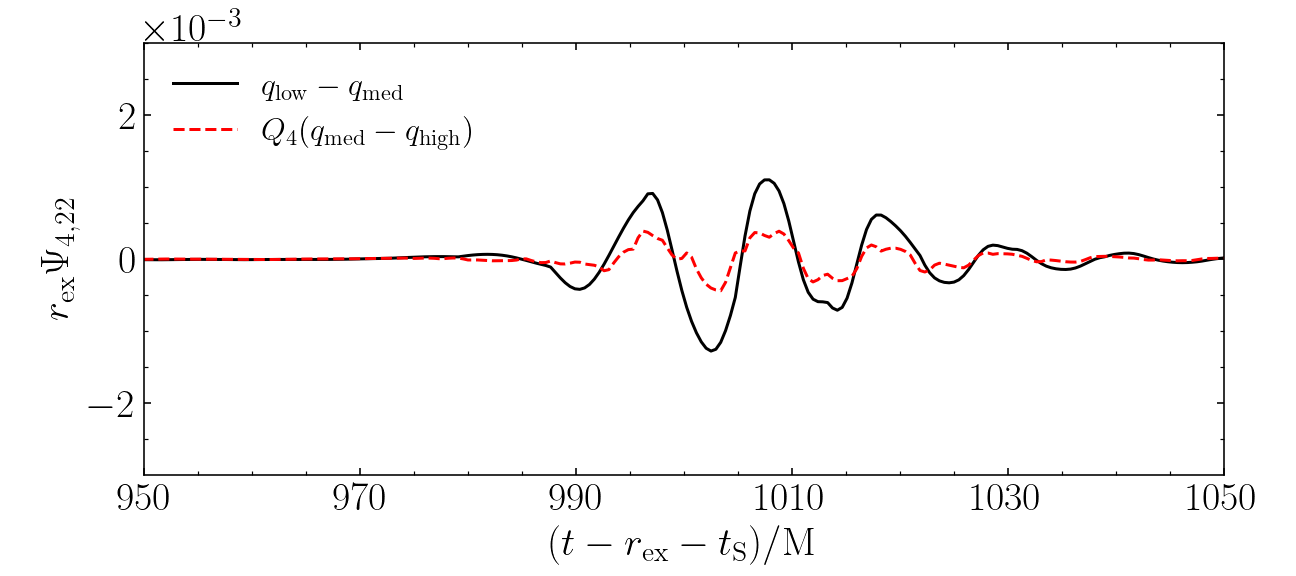}
    \caption{\label{fig:ZWConvergence_Psi4,22}
    Convergence plot of the gravitational radiation in a zoom-whirl simulation from the Xp0P24 series.
    \underline{Top:} Convergence test for radiation emitted during the \bh{s'} first encounter.
    \underline{Bottom:} Convergence tests for the radiation emitted during the \bh{s'} merger. The simulation times are shifted
    by $t_{\rm S}$ 
    to align at the peak of the low resolution run.
    }
    \end{center}
\end{figure}

The convergence test for the dominant mode of the gravitational radiation is displayed in Fig.~\ref{fig:ZWConvergence_Psi4,22}.
The convergence plot of the first encounter is displayed in the top panel, and the convergence plot of the merger is displayed in the bottom panel.
The peak in both panels approximately coincides with the peak of the waveform.
We display the difference between the low and medium resolutions along with the difference between the medium and high resolutions multiplied by the convergence factor, $Q_{\rm 4}=0.634$, indicating fourth order convergence.
The percent errors at the waveform peaks are about $0.14 \%$ and $1.6 \%$ during the first encounter and merger, respectively.

\begin{figure}[htbp!]
    \begin{center}

    \includegraphics[height=0.833\columnwidth]{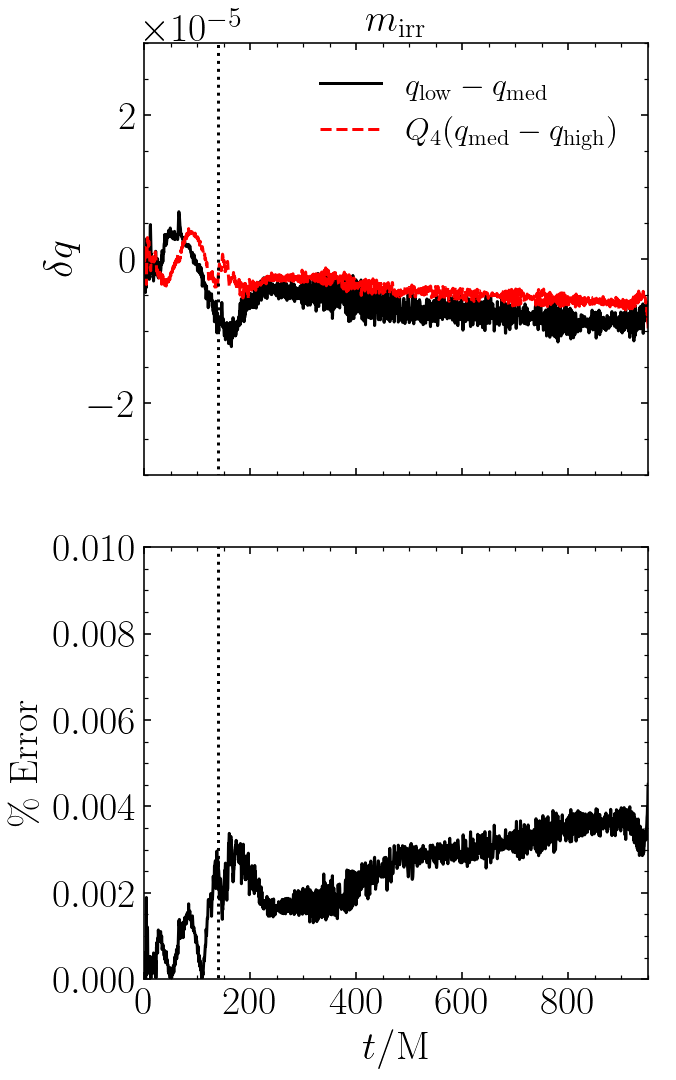} \hspace{-0.022\columnwidth}
    \includegraphics[height=0.833\columnwidth]{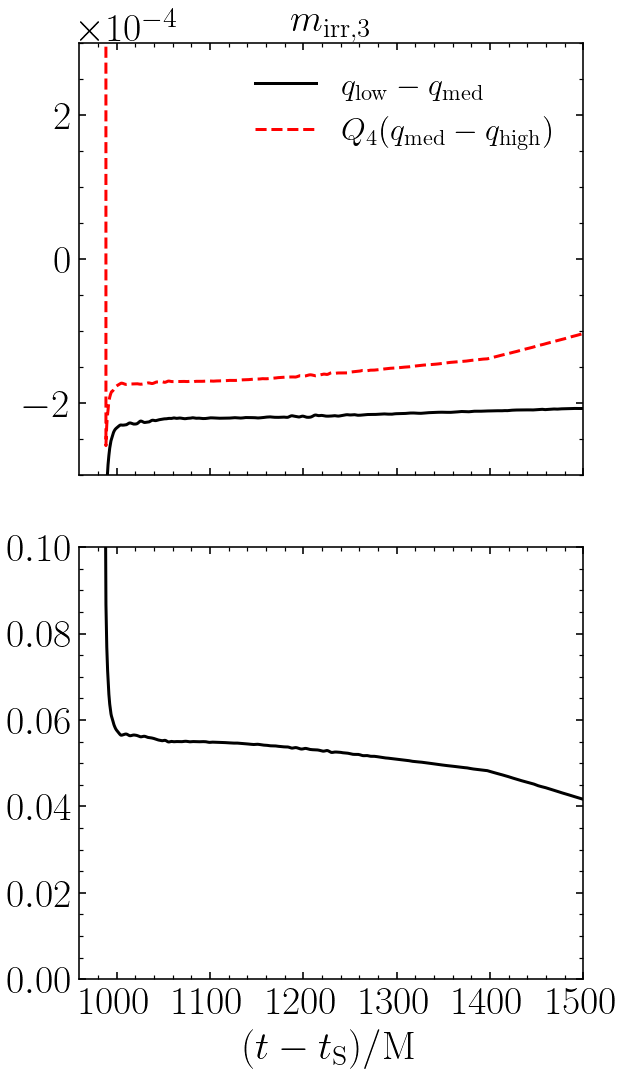}

    \caption{\label{fig:ZWConvergence_IrreducibleMass} Convergence plot (top panels) and percent error (bottom panels) of the irreducible mass in a zoom-whirl simulation from the Xp0P24 series. \underline{Left:} Irreducible mass of one of the \bh{s} prior to merger. The dotted line denotes the time of closest encounter. \underline{Right:} Irreducible mass of the remnant \bh{} after the merger. Here, time is shifted by $t_{\rm S}$ such that the time of merger coincides with the time in the $dx_{\rm low}=1.0$ simulation.
}
    \end{center}
\end{figure}

The convergence and error analysis of the irreducible mass is displayed in Fig.~\ref{fig:ZWConvergence_IrreducibleMass}.
The panel on the top left shows the convergence test for the irreducible mass of one of the 
\bh{s} prior to merger.
The panel on the top right shows the convergence test for the irreducible mass of the remnant \bh{}. The time of the encounter is denoted by a dotted line in the left panels. We find 4th order convergence. The bottom panels show the corresponding percent errors computed as a function of time. We find a percent error of about $0.002\%$ before the first encounter and about $0.004\%$ after the first encounter. The percent error of the remnant \bh{'s} mass is around $0.06 \%$.

\begin{figure}[htbp!]
    \begin{center}

    \includegraphics[height=0.833\columnwidth]{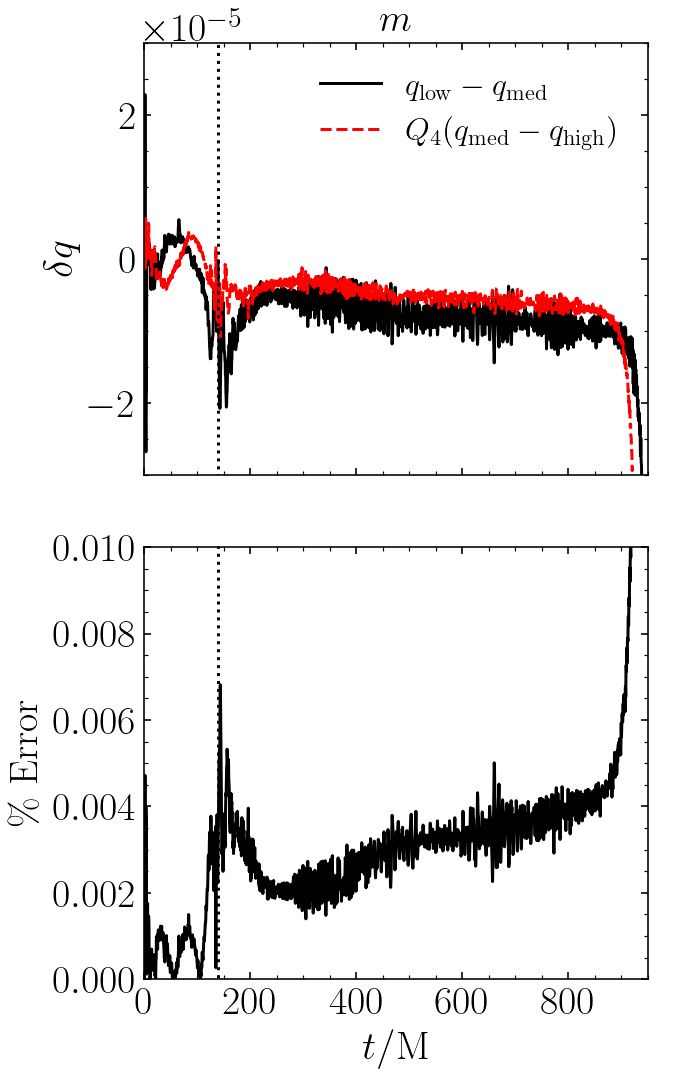} \hspace{-0.022\columnwidth}
    \includegraphics[height=0.833\columnwidth]{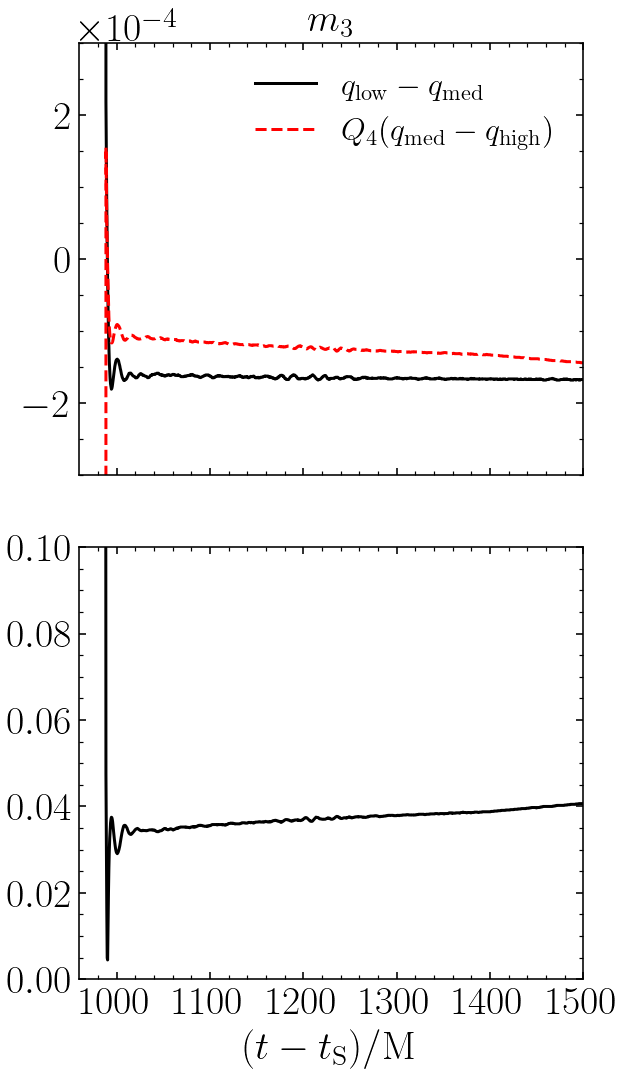}

    \caption{\label{fig:ZWConvergence_ChrisMass} Convergence plot (top panels) and percent error (bottom panels) of the \bh{} mass in a zoom-whirl simulation from the Xp0P24 series. \underline{Left:} \bh{} mass of one of the \bh{s} prior to merger. The dotted line denotes the time of closest encounter. \underline{Right:} \bh{} mass of the remnant \bh{} after the merger. Here, time is shifted by $t_{\rm S}$ such that the time of merger coincides with the time in the $dx_{\rm low}=1.0$ simulation.}
    \end{center}
\end{figure}

The convergence and error analysis of the \bh{} mass is displayed in Fig.~\ref{fig:ZWConvergence_ChrisMass}. The panel on the top left shows the convergence test for the \bh{} mass of one of the \bh{s} prior to merger. The panel on the top right shows the convergence test for the \bh{} mass of the remnant \bh{}. The time of the encounter is denoted by a dotted line in the left panels. We find fourth order convergence. The bottom panels show the corresponding percent errors computed as a function of time. We find a percent error of about $0.001\%$ before the first encounter and about $0.005\%$ after the first encounter. The percent error of the remnant \bh{'s} mass is around $0.04 \%$.

\begin{figure}[htbp!]
    \begin{center}
    \includegraphics[height=0.833\columnwidth]{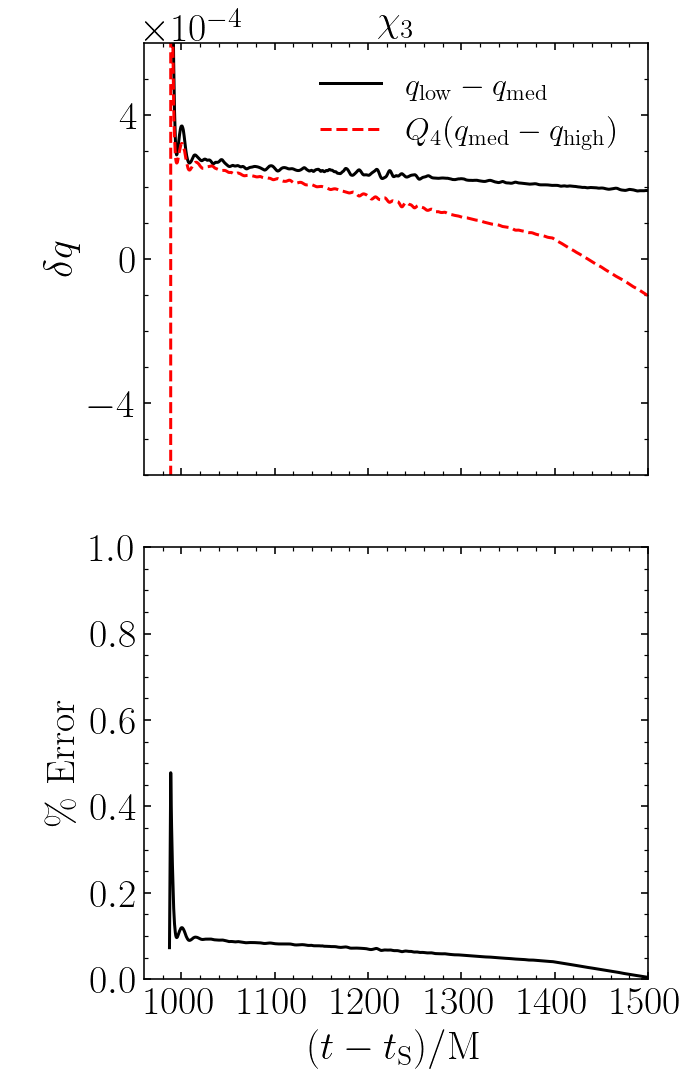} \hspace{-0.022\columnwidth}
    \includegraphics[height=0.833\columnwidth]{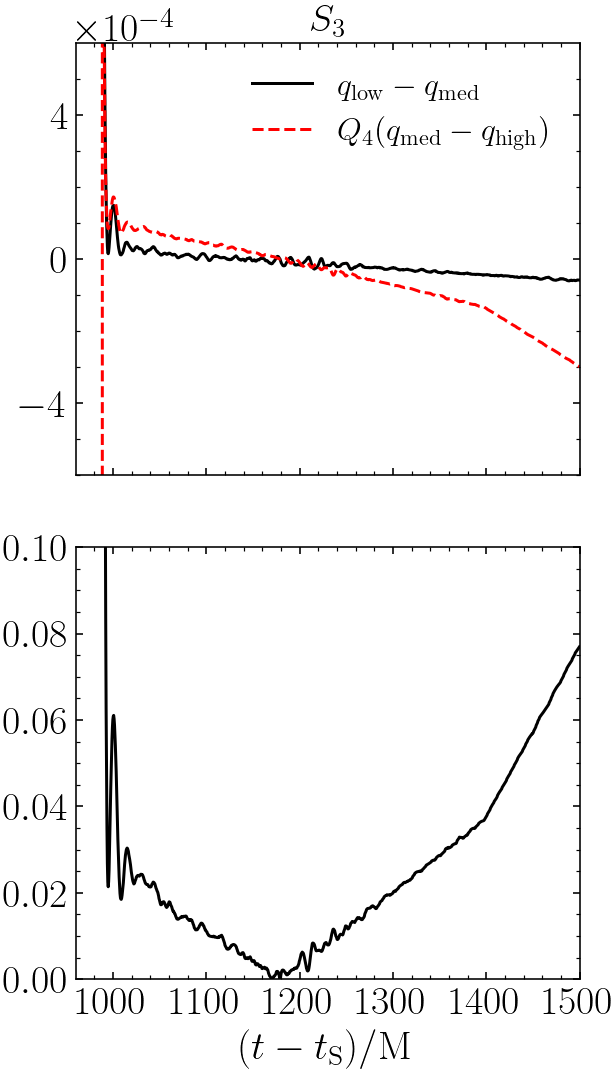}
    \caption{\label{fig:ZWConvergence_spin&AngMomMerge} Convergence plot (top panels) and percent error (bottom panels) of the spin and \bh{} angular momentum of the remnant \bh{} in a zoom-whirl simulation from the Xp0P24 series. \underline{Left:} Spin of the remnant \bh{} after the merger. \underline{Right:} \bh{} angular momentum of the remnant \bh{} after the merger. Here, the time is shifted by $t_{\rm S}$, such that the time of merger coincides with the time in the $dx_{\rm low}=1.0$ simulation.
    }
    \end{center}
\end{figure}

The convergence and error analysis of the remnant \bh{}'s spin is shown on the left of Fig.~\ref{fig:ZWConvergence_spin&AngMomMerge}.
The top panel shows the convergence test for the remnant's spin,
and we find fourth order convergence.
The bottom panel shows the corresponding percent error 
as a function of time.
The relative error is ill-defined prior to the encounter as the \bh{s} are initially non-spinning.
During their encounter, the \bh{s} spin up to $\chi\sim0.2$, so the relative error is well defined
after the encounter.
We find a percent error of about $5\%$ after the first encounter. The percent error of the remnant \bh{} stabilizes to around $0.1 \%$.

The convergence and error analysis of the remnant \bh{}'s angular momentum is displayed on the right of Fig.~\ref{fig:ZWConvergence_spin&AngMomMerge}. The top panel shows the convergence test for the \bh{} angular momentum of the remnant \bh{}. We find fourth order convergence. The bottom panel shows the corresponding percent error computed as a function of time. Although the percent error of the \bh{} angular momentum is ill-defined before the first encounter, we find a percent error of about $5\%$ after the first encounter. The percent error of the remnant \bh{} is below $0.1\%$.

\subsection{Scattering of spinning black holes} \label{sec:convergence_tests_Spin}

We seek to understand how the error in the simulation suites depends on the initial spin and incident angle.
We analyze this dependence 
by
testing two scattering simulations with initial spin magnitude $|\chi_{\rm i}|=0.7$, such that one case has an incident angle far from the threshold value
and the other case has an incident angle close to the threshold value.

\subsubsection{Far From Threshold: $\theta_{\rm th}<\theta=0.05685$, $\chi_{\rm i}=0.7$} \label{sec:convergence_tests_Spin_FFT}

Here we present the convergence test for the scattering simulation from the Xp7P49 series with incident angle $\theta=0.05685$, positive initial spin $\chi_{\rm i}=0.7$, and initial momentum $|\vec{P}_{\rm i}|=0.490\mathrm{M}$. The incident angle is selected to be large compared to the threshold angle, $\theta_{\rm th}=0.04250$. This test provides us with insight into how the simulations behave for large incident angles.

\begin{figure}[htbp!]
    \begin{center}
    \includegraphics[width=1\columnwidth]{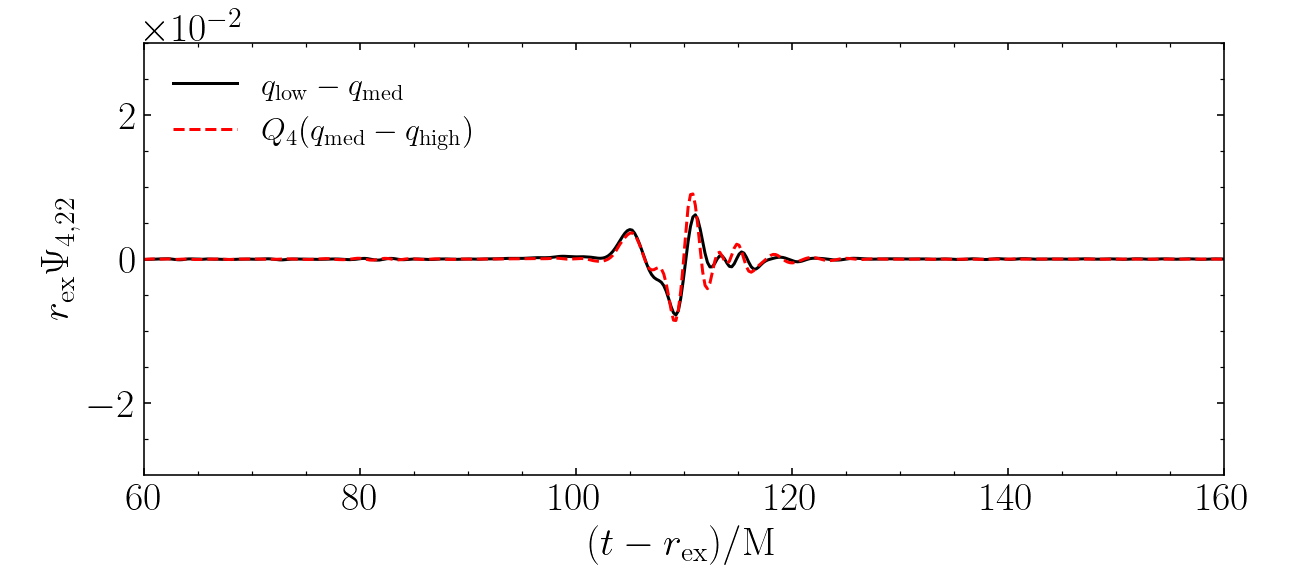}
    \caption{\label{fig:SpinConvergence_Psi4-22_The005685_spinp07}
   Convergence plot of the gravitational radiation in a scattering simulation from the Xp7P49 series with incident angle far from the threshold angle. The plot is centered on the pulse of radiation emitted during the encounter between the \bh{s}.
    }
    \end{center}
\end{figure}

The convergence test for the dominant mode of the gravitational radiation is displayed in Fig. \ref{fig:SpinConvergence_Psi4-22_The005685_spinp07}. The peak in Fig. \ref{fig:SpinConvergence_Psi4-22_The005685_spinp07} roughly coincides with the peak of the waveform emitted during the \bh{s}' encounter. We display the difference between the low and medium resolutions along with the difference between the medium and high resolutions multiplied by the convergence factor, $Q_{\rm 4}=0.634$, indicating fourth order convergence. The percent error at the waveform peak is about $11.8 \%$.

\begin{figure}[htbp!]
    \begin{center}
    \includegraphics[height=0.833\columnwidth]{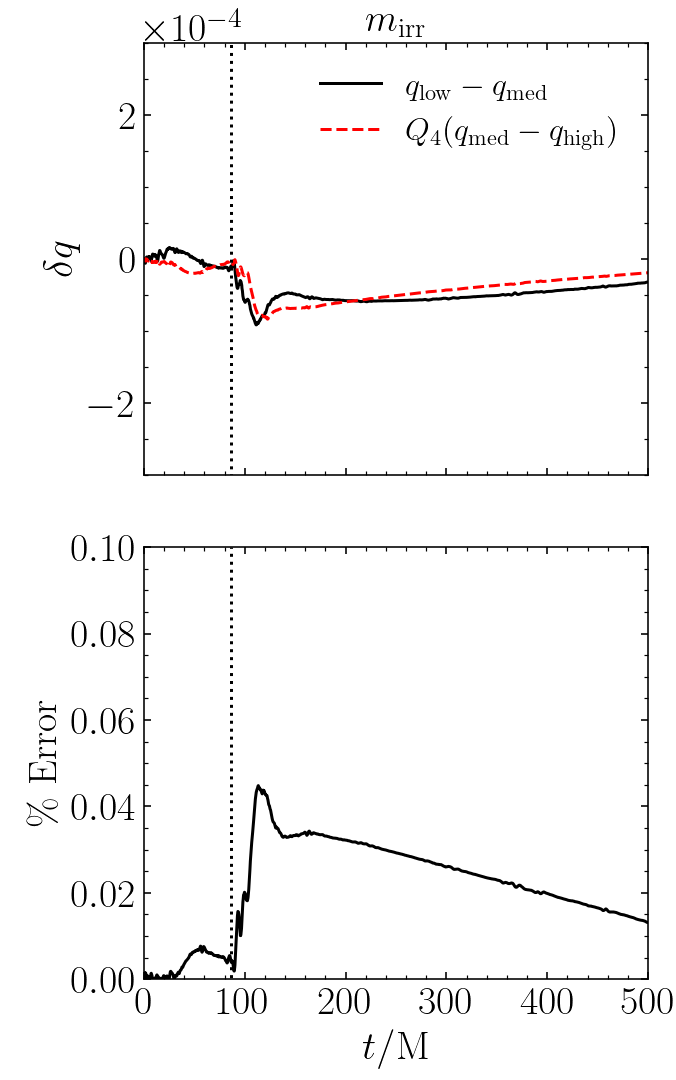} \hspace{-0.022\columnwidth}
    \includegraphics[height=0.833\columnwidth]{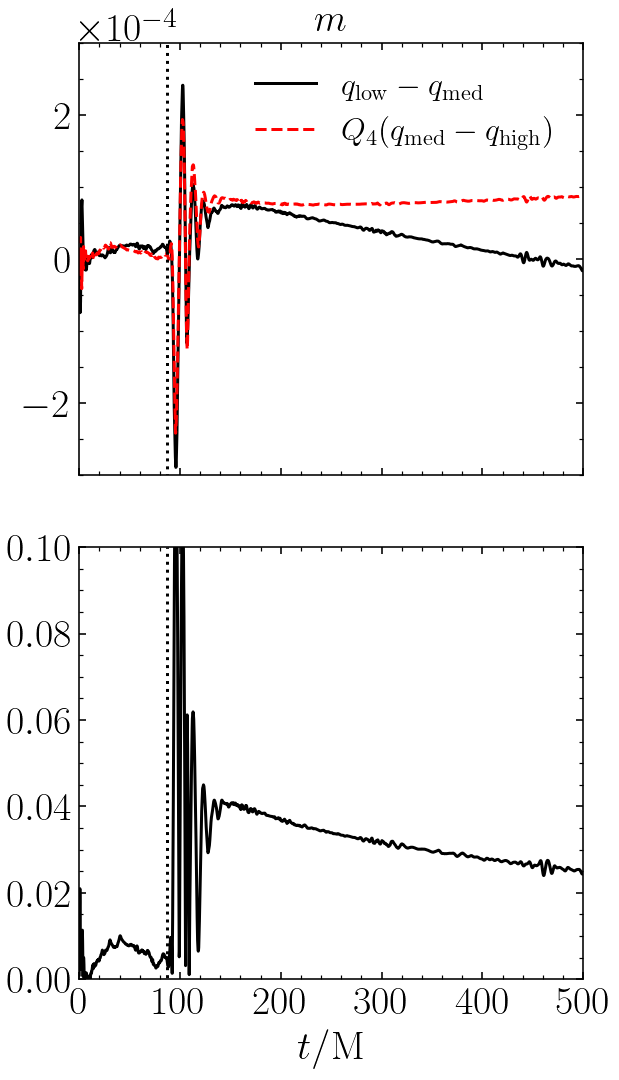}
    \caption{\label{fig:SpinConvergence_The005685_spinp07_Mass&Mirr}
    Convergence plot (top panels) and percent error (bottom panels) of the irreducible mass and \bh{} mass in a scattering simulation from the Xp7P49 series with incident angle far from the threshold angle. \underline{Left:} Irreducible mass of one of the \bh{s}. \underline{Right:} \bh{} mass of one of the \bh{s}. The dotted line denotes the time of closest encounter.
    }
    \end{center}
\end{figure}

The convergence and error analysis of the irreducible mass is displayed on the left of Fig.~\ref{fig:SpinConvergence_The005685_spinp07_Mass&Mirr}. The top panel shows the convergence test for the irreducible mass of one of the \bh{s}. The time of the encounter is denoted by a dotted line. We find fourth order convergence. The bottom panel shows the corresponding percent error computed as a function of time. We find a percent error of about $0.01 \%$ before the encounter and below $0.04\%$ after the encounter.

The convergence and error analysis of the \bh{} mass is displayed on the right of Fig.~\ref{fig:SpinConvergence_The005685_spinp07_Mass&Mirr}. The top panel shows the convergence test for the \bh{} mass of one of the \bh{s}. The time of the encounter is denoted by a dotted line. We find fourth order convergence. The bottom panel shows the corresponding percent error computed as a function of time. We find a percent error of about $0.01 \%$ before the encounter and below $0.04\%$ after the encounter.

\begin{figure}[htbp!]
    \begin{center}
    \includegraphics[height=0.833\columnwidth]{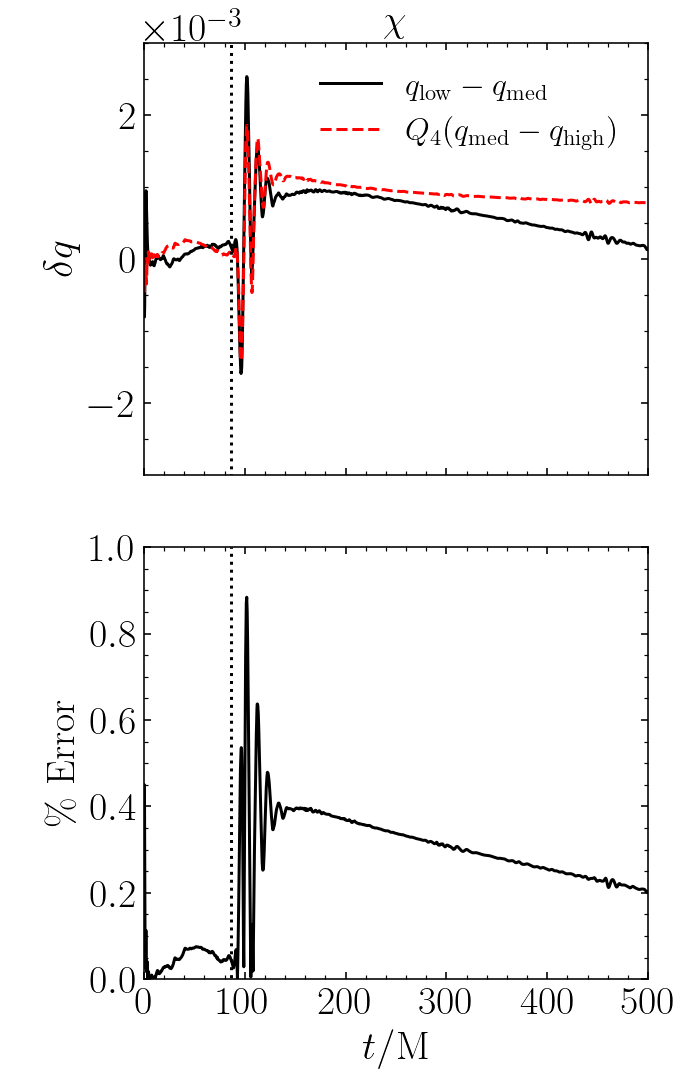} \hspace{-0.022\columnwidth}
    \includegraphics[height=0.833\columnwidth]{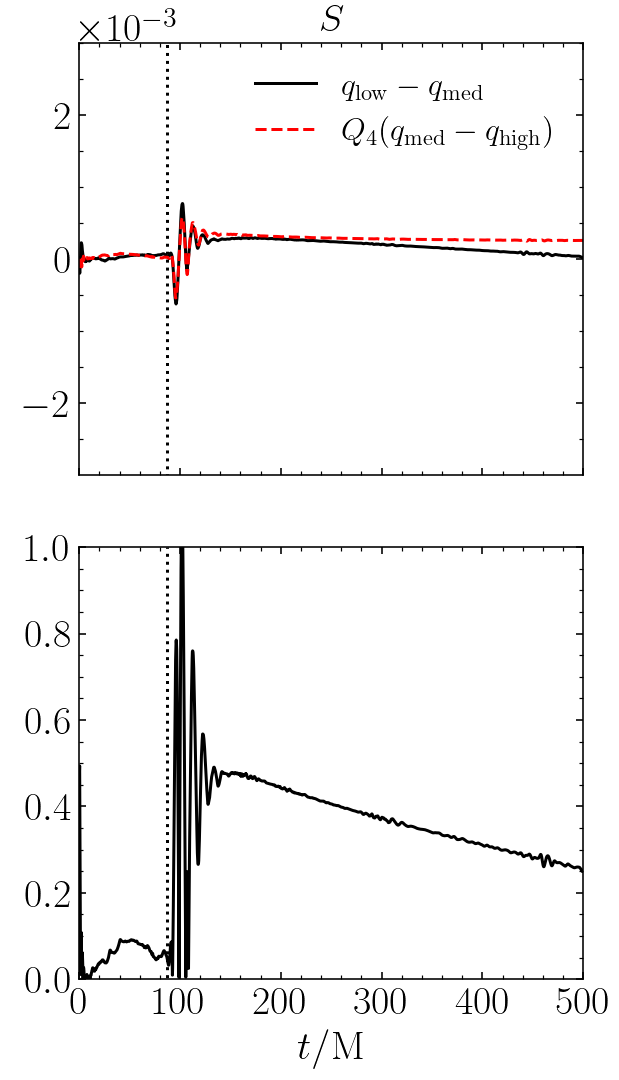}
    \caption{\label{fig:SpinConvergence_The005685_spinp07_spin&angmom}
    Convergence plot (top panels) and percent error (bottom panels) of the spin and \bh{} angular momentum in a scattering simulation from the Xp7P49 series with incident angle far from the threshold angle. \underline{Left:} Spin of one of the \bh{s}. \underline{Right:} \bh{} angular momentum of one of the \bh{s}. The dotted line denotes the time of closest encounter.
    }
    \end{center}
\end{figure}

The convergence and error analysis of the spin is displayed on the left of Fig. \ref{fig:SpinConvergence_The005685_spinp07_spin&angmom}. The top panel shows the convergence test for the spin of one of the \bh{s}. The time of the encounter is denoted by a dotted line. We find fourth order convergence. The bottom panel shows the corresponding percent error computed as a function of time. We find a percent error of about $0.1\%$ before the encounter and below $0.4 \%$ after the encounter.

The convergence and error analysis of the \bh{} angular momentum is displayed on the right of Fig. \ref{fig:SpinConvergence_The005685_spinp07_spin&angmom}. The top panel shows the convergence test for the \bh{} angular momentum of one of the \bh{s}. The time of the encounter is denoted by a dotted line. We find fourth order convergence. The bottom panel shows the corresponding percent error computed as a function of time. We find a percent error of about $0.1\%$ before the encounter and below $0.5\%$ after the encounter.

\subsubsection{Near Threshold: $\theta=\theta_{\rm th}=0.05685$, $\chi_{\rm i}=-0.7$} \label{sec:convergence_tests_Spin_NT}

Here we present the convergence test for the scattering simulation from the Xm7P49 series with incident angle $\theta=0.05685$, negative initial spin $\chi_{\rm i}=-0.7$, and initial momentum $|\vec{P}_{\rm i}|=0.490\mathrm{M}$. The incident angle is selected to be equal to the threshold angle. This test provides us with insight into how the simulations behave for small angles near the cutoff between scatterings and mergers. 

\begin{figure}[htbp!]
    \begin{center}
    \includegraphics[width=1\columnwidth]{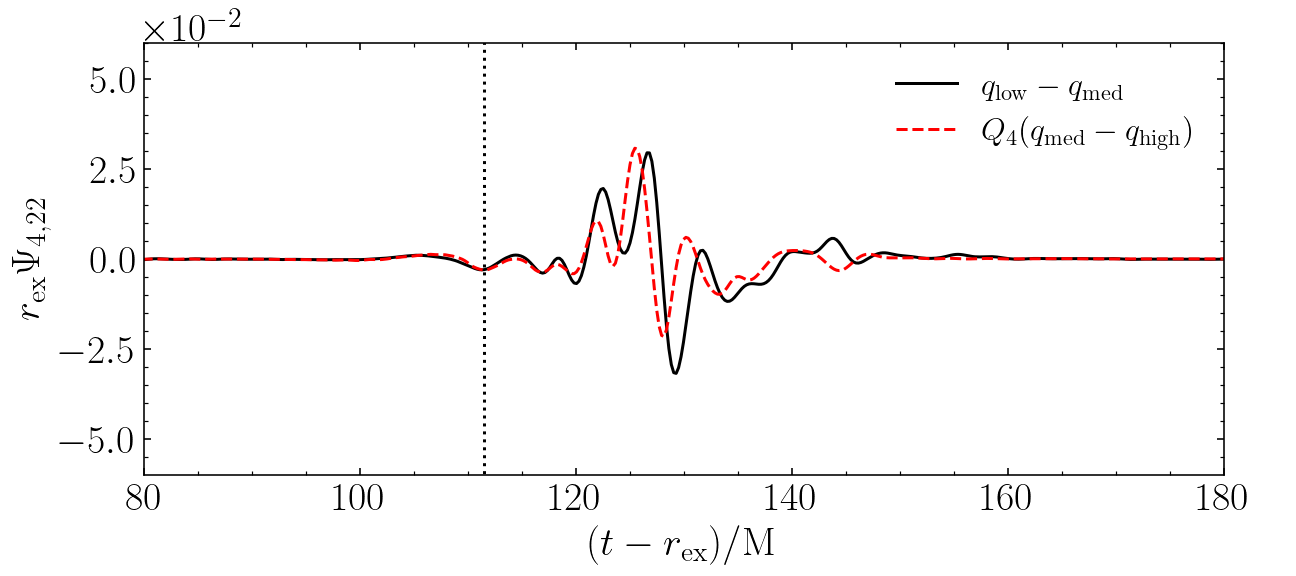}
    \caption{\label{fig:SpinConvergence_Psi4-22_The005685_spinm07}
    Convergence plot of the gravitational radiation in a scattering simulation from the Xm7P49 series with incident angle near to the threshold value. The plot is centered around the time of scattering. The time at which the waveform reaches its peak value is marked by a dotted line.
    }
    \end{center}
\end{figure}

The convergence test for the dominant mode of the gravitational radiation is displayed in Fig.~\ref{fig:SpinConvergence_Psi4-22_The005685_spinm07}. We display the difference between the low and medium resolutions along with the difference between the medium and high resolutions multiplied by the convergence factor, $Q_{\rm 4}=0.634$, indicating fourth order convergence. The percent error at the waveform peak is approximately $3.0\%$. However, the different resolutions deviate from one another more substantially in a small region towards the end of the waveform. Consequently, the peak of the plot in Fig.~\ref{fig:SpinConvergence_Psi4-22_The005685_spinm07} occurs some time after the peak of the waveform, which we denote with a vertical dotted line.

\begin{figure}[htbp!]
    \begin{center}
    \includegraphics[height=0.833\columnwidth]{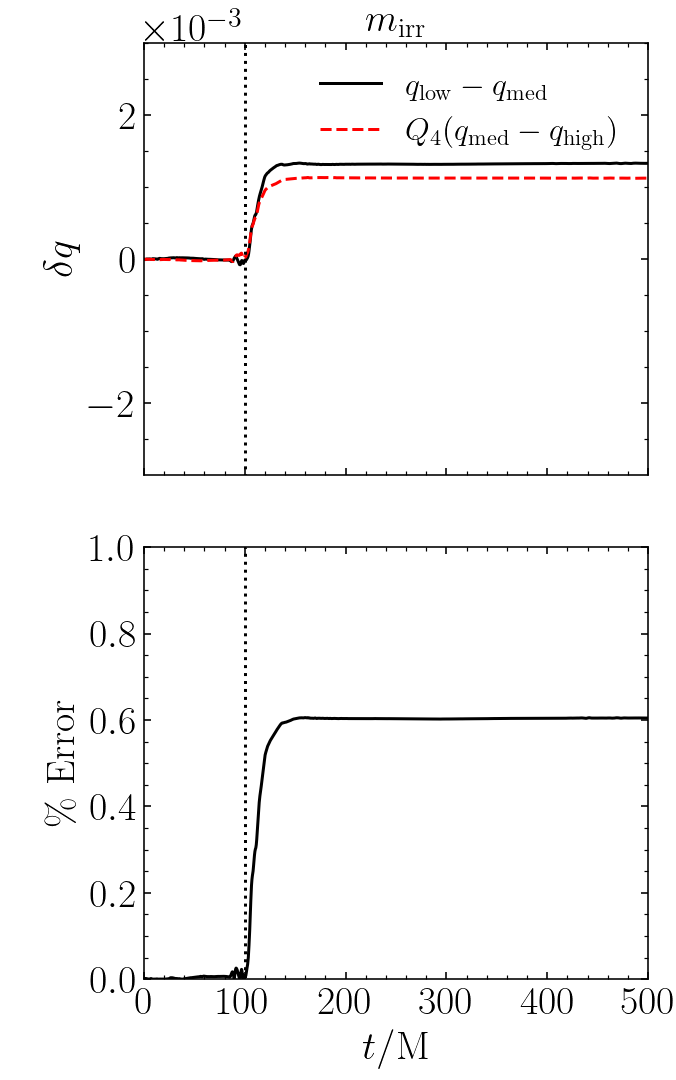} \hspace{-0.022\columnwidth}
    \includegraphics[height=0.833\columnwidth]{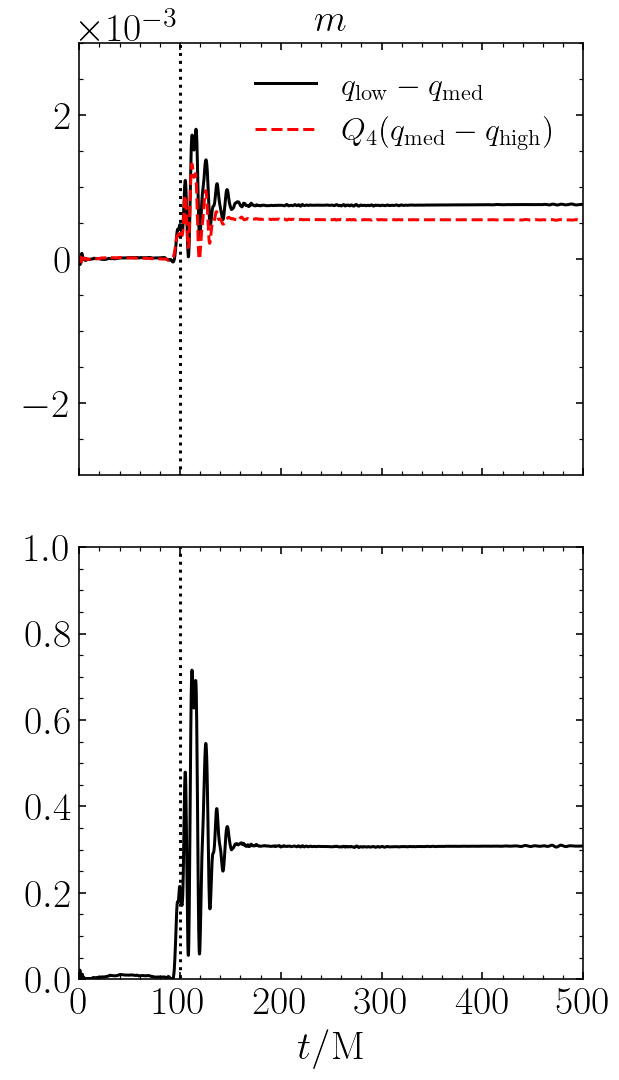}

    \caption{\label{fig:SpinConvergence_The005685_spinm07_Mass&Mirr}
    Convergence plots (top panels) and percent error (bottom panels) of the irreducible mass and \bh{} mass in a scattering simulation from the Xm7P49 series with incident angle near to the threshold value. \underline{Left:} Irreducible mass of one of the \bh{s}. \underline{Right:} \bh{} mass of one of the \bh{s}. The dotted line denotes the time of closest encounter.
    }
    \end{center}
\end{figure}

The convergence and error analysis of the irreducible mass is displayed on the left of Fig.~\ref{fig:SpinConvergence_The005685_spinm07_Mass&Mirr}. The top panel shows the convergence test for the irreducible mass of one of the \bh{s}. The time of the encounter is denoted by a dotted line. We find fourth order convergence. The bottom panel shows the corresponding percent error computed as a function of time. We find a percent error of about $0.01\%$ before the encounter and about $0.6\%$ after the encounter.

The convergence and error analysis of the \bh{} mass is displayed on the right of Fig.~\ref{fig:SpinConvergence_The005685_spinm07_Mass&Mirr}. The top panel shows the convergence test for the \bh{} mass of one of the \bh{s}. The time of the encounter is denoted with a dotted line. We find fourth order convergence. The bottom panel shows the corresponding percent error computed as a function of time. We find a percent error of about $0.01 \%$ before the encounter and about $0.3\%$ after the encounter.

\begin{figure}[htbp!]
    \begin{center}
    \includegraphics[height=0.833\columnwidth]{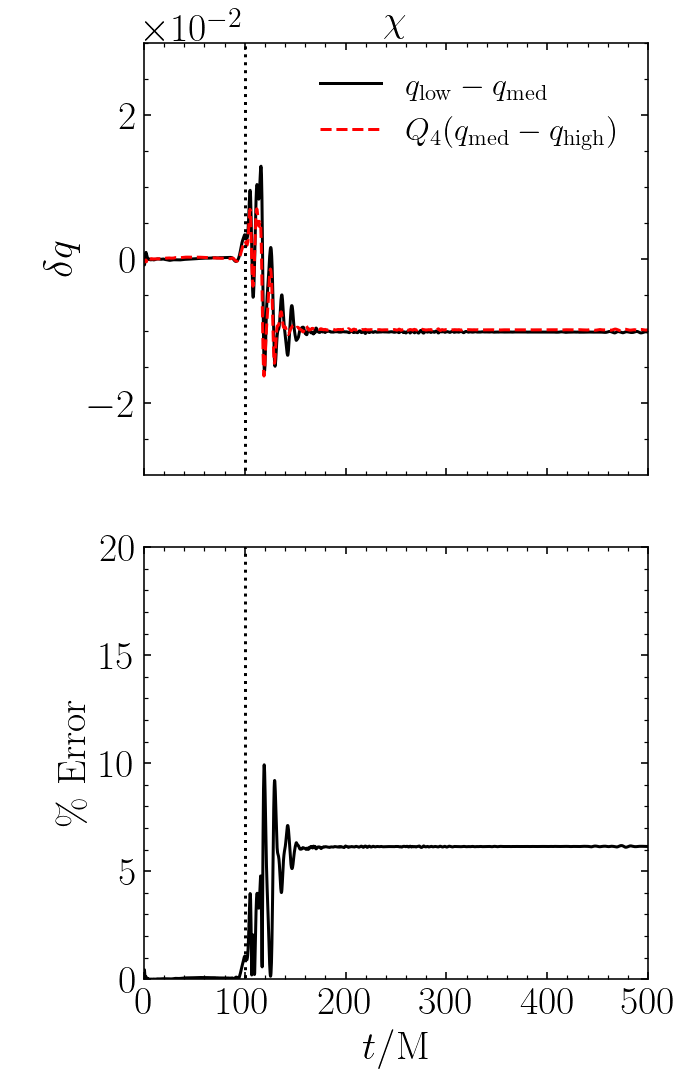} \hspace{-0.022\columnwidth}
    \includegraphics[height=0.833\columnwidth]{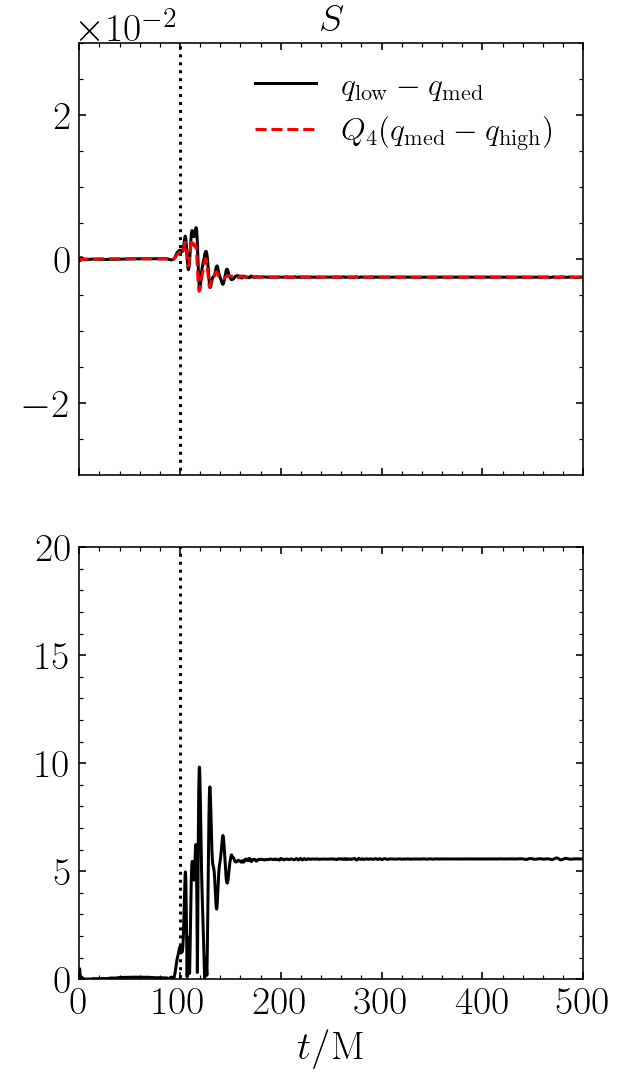}
    \caption{\label{fig:SpinConvergence_The005685_spinm07_spin&angmom}
    Convergence plot (top panels) and percent error (bottom panels) of the spin and \bh{} angular momentum in a scattering simulation from the Xm7P49 series with incident angle near to the threshold value. \underline{Left:} Spin of one of the \bh{s}. \underline{Right:} \bh{} angular momentum of one of the \bh{s}. The dotted line denotes the time of closest encounter.
    }
    \end{center}
\end{figure}

The convergence and error analysis of the spin is displayed on the left of Fig.~\ref{fig:SpinConvergence_The005685_spinm07_spin&angmom}. The top panel shows the convergence test for the spin of one of the \bh{s}. The time of the encounter is denoted by a dotted line. We find fourth order convergence. The bottom panel shows the corresponding percent error computed as a function of time. We find a percent error of about $0.1 \%$ before the encounter and about $6\%$ after the encounter.

The convergence and error analysis of the \bh{} angular momentum is displayed on the right of Fig.~\ref{fig:SpinConvergence_The005685_spinm07_spin&angmom}. The top panel shows the convergence test for the \bh{} angular momentum of one of the \bh{s}. The time of the encounter is denoted by a dotted line. We find fourth order convergence. The bottom panel shows the corresponding percent error computed as a function of time. We find a percent error of about $0.1\%$ before the encounter and about $6\%$ after the encounter.

\subsection{Uncertainty Estimates} \label{sec:convergence_tests_Uncertainty}

From the above tests, we estimate uncertainties on data reported in the main text.
Here we focus on the changes observed in the \bh{} parameters, and we refer to Appendix~\ref{sec:convergence_tests_Spin} for a summary of the percent error in the gravitational waveforms.
Typically, we find uncertainties much smaller than the changes in parameters observed for scattering \bh{s} with low initial spins and large incident angles. However, we find that uncertainties can be larger than the changes in parameters observed in systems with positive initial spins $\chi_{\rm i}>0.2$
and incident angles close to the threshold value.
This finding mainly affects simulations run with small initial momenta of $|\vec{P}_{\rm i}|\le0.245\mathrm{M}$, where the observed changes in parameters are smallest.

\subsubsection{Black Hole Mass and Irreducible Mass} \label{sec:convergence_tests_Uncertainty_mass&irrmass}

Here we discuss the uncertainty in the relative changes of the \bh{} mass and irreducible mass shown in Figs.~\ref{fig:MassvAngle} and~\ref{fig:MassvAngleBoosts}.
We find that the percent error in the \bh{} mass and irreducible mass after an encounter is larger than the percent error before an encounter.
Furthermore, we find that the changes in the \bh{} mass and irreducible mass due to the encounter are small relative to their initial values. Therefore, we take the final (i.e. post-encounter) percent error, $\% \mathrm{Error}_{\rm f}$, as an estimate of the uncertainty in the relative change of the \bh{} mass and irreducible mass,
\begin{equation} \label{eq:mass_uncertainty}
    \Delta \left( \frac{\delta m_{\rm (irr)}}{m_{\rm (irr),i}} \right) \simeq \frac{\%\mathrm{Error}_{\rm f}}{100} \,,
\end{equation}
where the percent error is given in Eq.~\eqref{eq:Error}.

\textbf{\underline{Irreducible Mass:}} The Xp0P24 test and Xp7P49 test have low post-encounter percent errors of $\lesssim0.04\%$. One can see that $0.04\%$ is far below the relative changes in the \bh{} mass reported in Figs.~\ref{fig:MassvAngle} and~\ref{fig:MassvAngleBoosts},
except for small initial momenta $|\vec{P}_{\rm i}|\leq0.1225\mathrm{M}$.
Thus, the uncertainty is negligible for high initial spins at large incident angles and low initial spins at any incident angle.

The Xm7P49 test has a post-encounter percent error of $0.6\%$. This error is greater than the relative changes in the irreducible mass that we find at small incident angles in Fig.~\ref{fig:MassvAngle} for systems with initial momentum $|\vec{P}_{\rm i}|=0.245\mathrm{M}$ and positive initial spin. 
Hence, the relative changes in the irreducible mass found in systems with low initial momenta,
positive initial spins, and small incident angles are consistent with zero within numerical error.

\textbf{\underline{Black Hole Mass:}} The percent errors that we find for the \bh{} mass are comparable to those that we find for the irreducible mass. Namely, the post-encounter percent errors are small for the Xp0P24 and Xp7P49 tests, with values $<0.04\%$. This error is at most comparable to some of the smaller changes in the \bh{} mass reported in Fig.~\ref{fig:MassvAngle} at larger incident angles for initial momentum $|\vec{P}_{\rm i}|=0.245\mathrm{M}$, and to the relative changes in \bh{} mass reported for initial momenta $|\vec{P}_{\rm i}|\leq0.1225\mathrm{M}$ in Fig.~\ref{fig:MassvAngleBoosts}. Thus, the uncertainty should be negligible for high initial spins at large incident angles and low initial spins at any incident angle, especially for higher initial momenta.

The Xm7P49 test has a post-encounter error of $0.3\%$. This error is greater than the relative changes in \bh{} mass that we find at small incident angles in Fig.~\ref{fig:MassvAngle} for systems with initial momentum $|\vec{P}_{\rm i}|=0.245\mathrm{M}$ and positive initial spin. Consequently, the relative changes in \bh{} mass found in systems with low initial momentum, positive initial spin, and small incident angles are consistent with zero within numerical error.

\subsubsection{Spin and Black-Hole Angular Momentum} \label{sec:convergence_tests_Uncertainty_spin&angmom}

Here we discuss the uncertainty of the changes in the spin and \bh{} angular momentum shown in Figs.~\ref{fig:SpinvAngle} and~\ref{fig:SpinvAngleBoosts}. We find that the percent error in the spin and \bh{} angular momentum is larger after an encounter than before. Therefore, the percent error in the change in spin and \bh{} angular momentum should be similar to the final (i.e. post-encounter) error, $\%\mathrm{Error}_{\rm f}$, in those quantities.
To estimate the uncertainty, we then need to multiply the error by the final absolute value of the quantity. The initial values of both quantities tend to be easier to compute and are generally close to or greater in magnitude than the final quantities. Therefore, we use the initial values as estimates for the final values when computing an uncertainty. In summary, we estimate,
\begin{subequations} \label{eq:spin_uncertainty}
\begin{align}
\Delta(\chi_{\rm f}-\chi_{\rm i}) & \simeq|\chi_{\rm i}|(\%\mathrm{Error}_{\rm f}/100) \, , \\
\Delta(S_{\rm f}-S_{\rm i})& \simeq m_{\rm i}^2|\chi_{\rm i}|(\%\mathrm{Error}_{\rm f}/100) \, ,
\end{align}
\end{subequations}
where the percent error is given in Eq.~\eqref{eq:Error}.

\textbf{\underline{Spin:}} We find that the Xp0P24 and Xm7P49 tests have post-encounter percent errors of about $5\%$ and $6\%$, respectively. Given that the former has an initial spin of $\chi_{\rm i}=0$ and the latter has an initial spin magnitude of $|\chi_{\rm i}|=0.7$, this similarity suggests that the near threshold error is largely independent of the initial spin. In the Xp7P49 test, we find a percent error of about $0.4\%$. These numbers suggest that the error tends to decline at larger incident angles.

\underline{Far From Threshold:} For an initial spin magnitude of $|\chi_{\rm i}|=0.7$, we can infer an uncertainty of $\Delta(\chi_{\rm f}-\chi_{\rm i})\simeq0.003$. This uncertainty is comparable to the changes in spin reported for initial spins $\chi_{\rm i} \geq 0.5$ and initial momenta $|\vec{P}_{\rm i}|\leq0.245\mathrm{M}$ at all incident angles (see left of Fig.~\ref{fig:SpinvAngle} and Fig.~\ref{fig:SpinvAngleBoosts}). Furthermore, the same appears to be true for initial spin $\chi_{\rm i}=0.5$ and initial momentum $|\vec{P}_{\rm i}|=0.490\mathrm{M}$ (see right of Fig.~\ref{fig:SpinvAngle}). These changes in spin are thus consistent with zero within numerical error. However, this uncertainty is notably lower than the spin-down that we observe for initial spin $\chi_{\rm i}=0.7$ systems at large angles in Fig.~\ref{fig:SpinvAngleBoosts}. This comparison tells us that the spin-down in these systems is a physical phenomena.

\underline{Near Threshold:} We find an uncertainty of about $\Delta(\chi_{\rm f}-\chi_{\rm i})\simeq0.04$ for initial spin magnitude $|\chi_{\rm i}|=0.7$. This uncertainty is larger than the observed changes in spin near the threshold angle in Fig.~\ref{fig:SpinvAngleBoosts} for all initial momenta. Therefore, the changes in spin reported in systems with initial spin $\chi_{\rm i}=0.7$ and small incident angles are consistent with zero within numerical error.

An initial spin magnitude of $|\chi_{\rm i}|=0.2$ gives an uncertainty of about $\Delta(\chi_{\rm f}-\chi_{\rm i})\simeq0.01$. This uncertainty is similar to the near threshold change in spin found for initial spin $\chi_{\rm i}=0.2$ and initial momentum $|\vec{P}_{\rm i}|=0.245\mathrm{M}$ in Fig.~\ref{fig:SpinvAngle}. Extrapolating from this observation, it is clear that the spin-up found in Fig.~\ref{fig:SpinvAngle} is greater than the corresponding uncertainty whenever $\chi_{\rm i}\leq0.2$ (including all negative initial spins).

\textbf{\underline{Black Hole Angular Momentum:}} The percent errors that we find for the \bh{} angular momentum are similar to those that we find for the (dimensionless) spin. We find that the Xp0P24 and Xm7P49 tests have post-encounter percent errors of $5\%$ and $6\%$, respectively. Thus, we again find that the near threshold error is largely independent of initial spin. The Xp7P49 test has a percent error of about $0.5\%$ suggesting that the error declines at larger incident angles.

\underline{Far From Threshold:} For an initial spin magnitude of $|\chi_{\rm i}|=0.7$, we can infer an uncertainty of about $\Delta(S_{\rm f}-S_{\rm i})\simeq0.001\mathrm{M}^2$. This uncertainty is comparable to the changes in \bh{} angular momentum reported for initial spin $\chi_{\rm i}\geq 0.5$ and initial momenta $|\vec{P}_{\rm i}|\leq0.245\mathrm{M}$ at all incident angles (see left of Fig.~\ref{fig:SpinvAngle} and Fig.~\ref{fig:SpinvAngleBoosts}). It also appears comparable to some of the data found at large incident angles for initial spin $\chi_{\rm i}=0.7$ and initial momentum $|\vec{P}_{\rm i}|=0.490\mathrm{M}$ in Figs.~\ref{fig:SpinvAngle} and~\ref{fig:SpinvAngleBoosts}. These changes in \bh{} angular momentum are therefore consistent with zero within numerical error.

\underline{Near Threshold:} We estimate an uncertainty of about $\Delta(S_{\rm f}-S_{\rm i})\simeq0.01\mathrm{M}^2$ for initial spin magnitude $|\chi_{\rm i}|=0.7$. This uncertainty is greater than the changes in \bh{} angular momentum reported at small incident angles for initial momenta $|\vec{P}_{\rm i}|\leq0.3675\mathrm{M}$ in Fig.~\ref{fig:SpinvAngleBoosts}. This tells us that the apparent negative changes in \bh{} angular momentum are consistent with zero within numerical error. However, this uncertainty is less than the increase in \bh{} angular momentum found at small incident angles for initial spin $\chi_{\rm i}\geq0.5$ and initial momenta $|\vec{P}_{\rm i}|\geq0.490\mathrm{M}$ (see right of Fig.~\ref{fig:SpinvAngle} and Fig.~\ref{fig:SpinvAngleBoosts}); these increases are physical (unlike the spin-up).

For initial spin magnitude $|\chi_{\rm i}|=0.2$, we estimate a near threshold uncertainty of about $\Delta(S_{\rm f}-S_{\rm i})\simeq0.003\mathrm{M}^2$. This uncertainty is similar to the change in \bh{} angular momentum found at small angles for initial spin $\chi_{\rm i}=0.2$ and initial momentum $|\vec{P}_{\rm i}|=0.245\mathrm{M}$ in Fig.~\ref{fig:SpinvAngle}. Extrapolating from this observation, it is clear that the data reported in Fig.~\ref{fig:SpinvAngle} is greater than the corresponding uncertainty whenever $\chi_{\rm i}\leq0.2$ (including all negative initial spins).

\bibliography{Refs_ScatteringGR.bib}
\end{document}